\documentclass[12pt]{article}

\usepackage{amsmath, amsthm, amsfonts}
\usepackage[latin1]{inputenc}

\usepackage{hyperref}           

\newcommand{\abs}[1]{\left\vert#1\right\vert}
\newcommand{\ap}[1]{\left\langle#1\right\rangle}
\newcommand{\norm}[1]{\left\Vert#1\right\Vert}

\def\RR{\mathbb{R}}

\newtheorem{thm}{Theorem}[section]

\newtheorem{lem}[thm]{Lemma}
\newtheorem{prp}[thm]{Proposition}

\theoremstyle{definition}
\newtheorem{dfn}[thm]{Definition}
\theoremstyle{remark}
\newtheorem{rem}[thm]{Remark}

\title{Regularity, asymptotic behavior and partial uniqueness for
  Smoluchowski's coagulation equation}

\author{Stéphane Mischler
  \thanks{Email: \texttt{mischler@ceremade.dauphine.fr}}
  \\
  CEREMADE, Univ. Paris-Dauphine
  \\
  Place du Maréchal de Lattre de Tassigny
  \\
  75775 Paris CEDEX 16, France
  \\
  José Alfredo Cañizo
  \thanks{Supported by the project MTM2008-06349-C03-03 DGI-MICINN
    (Spain). Email: \texttt{canizo@mat.uab.cat}}
  \\
  Departament de Matemàtiques
  \\
  Universitat Autònoma de Barcelona
  \\
  08193 Bellaterra (Barcelona), Spain }

\date{May 2009}

\begin{document}
\maketitle

\abstract{We consider Smoluchowski's equation with a homogeneous
  kernel of the form $a(x,y) = x^\alpha y ^\beta + x^\beta y^\alpha$
  with $-1 < \alpha \leq \beta < 1$ and $\lambda := \alpha + \beta \in
  (-1,1)$. We first show that self-similar solutions of this equation
  are infinitely differentiable and prove sharp results on the
  behavior of self-similar profiles at $y = 0$ in the case $\alpha <
  0$. We also give some partial uniqueness results for self-similar
  profiles: in the case $\alpha = 0$ we prove that two profiles with
  the same mass and moment of order $\lambda$ are necessarily equal,
  while in the case $\alpha < 0$ we prove that two profiles with the
  same moments of order $\alpha$ and $\beta$, and which are asymptotic
  at $y=0$, are equal. Our methods include a new representation of the
  coagulation operator, and estimates of its regularity using
  derivatives of fractional order.}

\tableofcontents

\section{Introduction}

\subsection{Smoluchowski's equation}

Smoluchowski's coagulation equation is a well-known model for
irreversible aggregation processes involving a set of particles which
can join to form groups of two or more of them, which we call
\emph{clusters}. Below we will briefly present the model, but we refer
the reader to the reviews \cite{A99,D72,MR2068589,citeulike:1134194}
for a more detailed background on the equation.

We study this equation in a continuous setting, meaning that the size
of a cluster may be any positive number $y \in (0,\infty)$. We are
interested in the time evolution of the density of clusters of each
possible size, given by a function $f = f(t,y)$ which depends on the
time $t$ and cluster size $y$. The \emph{mass} of $f$ at a given time
$t$ is given by its first moment $\int_0^\infty y \, f(t,y) \,dy$. The
continuous Smoluchowski equation reads:
\begin{equation}
  \label{eq:coag-eq}
  \partial_t f(t,y) = C(f(t),f(t))(y),
\end{equation}
where
\begin{equation}
  \label{eq:coag_operator}
  \begin{split}
    C(f,f)(y) := 
    &\frac{1}{2} \int_0^y a(z,y-z) f(z) f(y-z) \,dz
    \\
    &- \int_0^\infty a(z,y) f(z) f(y) \,dz,
  \end{split}
\end{equation}
where the \emph{coagulation kernel} $a = a(x,y)$ is a given
nonnegative symmetric function defined on $(0,+\infty) \times
(0,+\infty)$, which governs the time rate at which a cluster of size
$x$ and cluster of size $y$ aggregate. We write $C(f,f)$ to emphasize
that $C$ is a quadratic operator, and to agree with later discussions
in which we consider its associated symmetric bilinear form $C(f,g)$.

In many physical models the coagulation kernel $a$ is a homogeneous
function \cite{A99} in the sense that for some $\lambda \geq 0$ it
holds that
\begin{equation*}
  a(hx, hy) = h^\lambda a(x,y)
  \quad \text{ for } x,y,h > 0.
\end{equation*}
For our results we will take that as a hypothesis, and in fact we will
assume that $a$ is of the following form:
\begin{subequations}
  \label{eq:kernel-expression-precise}
  \begin{gather}
    \label{eq:kernel-expression-precise:form}
    a(x,y) := x^\alpha y^\beta + x^\beta y^\alpha \quad \text{ for }
    x, y > 0,
    \\
    \label{eq:kernel-expression-precise:conditions}
    -1 < \alpha \leq \beta < 1, \qquad \lambda := \alpha + \beta
    \in (-1,1).
  \end{gather}
\end{subequations}
In some places, we allow $a$ to be a finite linear combination of such
terms.

When the kernel is homogeneous, one may look for \emph{self-similar
  solutions}, also called \emph{scaling solutions}: solutions which
are given by a rescaling of some fixed function $g$ for all times. In
other words, a solution $f = f(t,y)$ to Smoluchowski's equation is
self-similar if there exists some nonnegative function $g$ such that
\begin{equation*}
  f(t,y) = q(t) \, g(p(t) y)
  \quad \text{ for all } t,y > 0
\end{equation*}
for some functions $q(t), p(t) > 0$. Such a function $g$ is called a
\emph{self-similar profile} (or \emph{scaling profile}). In this work
we will always consider self-similar solutions with finite mass. If
$a$ is homogeneous of degree $-1 < \lambda < 1$,
it is known (e.g., \cite{citeulike:1069567}) that for $f$ to be a
self-similar solution it must happen that $$q(t) = (t_0
+t)^{-\frac{2}{1-\lambda}} \quad \text{ and }\quad p(t) =
(t_0+t)^{-\frac{1}{1-\lambda}}$$ for some constant $t_0 > 0$, and that
$g$ satisfies the \emph{self-similar profile equation}:
\begin{equation}
  \label{eq:ss-profile-eq}
  2g + y \partial_y g
  + (1-\lambda) C(g,g)
  =
  0.
\end{equation}
The existence of self-similar profiles has been established in
\cite{citeulike:1172261,citeulike:1069567} for a general class of
coefficients which in particular cover the case of $a(x,y)$ satisfying
\eqref{eq:kernel-expression-precise}. It is expected (and has been
proved in some particular cases) that in general conditions solutions
exhibit a universal self-similar behavior in the long time, meaning
that they are eventually close to a self similar solution of the
equation. The conjecture that this behavior does in fact hold is
called the \emph{dynamical scaling hypothesis}, and a lot of effort
has been done to rigorously prove its validity. In particular, it is
expected that with ``regular'' initial conditions (such as compactly
supported data) solutions become eventually close to a self-similar
solution with finite mass. For the special cases $a(x,y) = 1$ or
$a(x,y) = x+y$ rather complete results have been proved
\cite{citeulike:1103487, citeulike:1103441,MR2139564}, but for a
general coefficient almost no information is available. It seems
likely that further study of the scaling profiles is needed before a
more complete understanding of dynamical scaling can be obtained, and
our results aim in this direction.

\subsection{Description of the main results}
\label{sec:main-results}

We first give a proof of the infinite differentiability of
self-similar profiles. This fact was already known in the case $\alpha
< 0$ \cite{citeulike:1069567}, and here we generalize this result to
include the $\alpha \geq 0$ case:
\begin{thm}
  \label{thm:local-regularity-intro}
  Assume that the coefficient $a$ is of the form
  \eqref{eq:kernel-expression-precise}, or is a finite linear
  combination of terms of that form, all with the same homogeneity
  degree $\lambda$. Then, any self-similar profile with finite mass
  for Smoluchowski's equation is infinitely differentiable on
  $(0,\infty)$.
\end{thm}
We remark that in the paper \cite{citeulike:1271620}, self-similar
profiles for $\alpha = 0$ were shown to be $\mathcal{C}^1$, and in
\cite[Th. 4.3]{citeulike:1069567} solutions for $\alpha \geq 0$ were
proved to have a $C^{0,\theta}$ Hölder regularity for any $0 \leq
\theta < 1-\lambda$.

Regarding the asymptotic behavior of scaling profiles at $0$ and
$\infty$, some estimates have been rigorously proved
\cite{citeulike:1172261,MR2114413,citeulike:1069567}, and are
consistent with the very complete previous formal calculations by van
Dongen and Ernst \cite{citeulike:1300131}. Exponential decay of
solutions as $y \to \infty$ was proved in \cite{citeulike:1069567,
  citeulike:1271620}, but the rate of decay is still far from the rate
expected by formal calculations in \cite{citeulike:1300131}. As for
the behavior at $y=0$, one has to separately treat the $\alpha = 0$,
$\alpha > 0$ and $\alpha < 0$ cases:

\begin{enumerate}
\item For $\alpha = 0$, a very precise result was given by Fournier
  and Laurençot in \cite{citeulike:1271620}, making rigorous the
  conclusions in \cite{citeulike:1300131}: a self-similar solution is
  asymptotic to a constant times $y^{-\tau}$ as $y \to 0$, for some $1
  < \tau < \min\{3/2, 1 + \lambda\}$. An intriguing property of this
  result is that in order to know the numerical value of $\tau$ one
  needs to compute the solution itself, as $\tau$ is given in terms of
  the $\lambda$-moment of the solution.
\item For $\alpha > 0$, the formal arguments in \cite{citeulike:1300131}
  suggest that a scaling profile $g$ should be asymptotic to a
  constant times $y^{-1-\lambda}$ when $y \to 0$. So far, it has been
  rigorously proved that all the moments of order $\sigma > \lambda$
  of $g$ are finite, while all moments of order $\sigma < \lambda$ are
  infinite.
\item For $\alpha < 0$, exponential decay of solutions at $y=0$ was
  proved in \cite{citeulike:1069567}. Our contribution here is a
  refinement of this behavior which coincides with the formal result
  in \cite{citeulike:1300131}, and which is needed in the proof of uniqueness
  for $\alpha < 0$.
\end{enumerate}
Our precise result for $\alpha < 0$ is the following:
\begin{thm}
  \label{thm:III-asymptotic-behavior-intro}
  Assume that the coagulation kernel is of the form
  \eqref{eq:kernel-expression-precise} with $\alpha < 0$. Then, if $g$
  is a (nonzero) self-similar profile for Smoluchowski's coagulation
  equation, it holds that
  \begin{equation}
    \label{eq:III-asymptotic-behavior}
    g(y)
    \sim
    K_0 e^{-\Lambda(y)}
  \end{equation}
  for some strictly positive constant $K_0$, where
  \begin{equation*}
    \Lambda(y) :=
    2 \log y
    - (1-\lambda) \left(
      \frac{M_\beta}{\alpha} y^\alpha
      +
      \frac{M_\alpha}{\beta} y^\beta
      \right),
  \end{equation*}
  and $M_\alpha$, $M_\beta$ are the moments of order $\alpha$ and
  $\beta$ of $g$. In addition, the function $y \mapsto g(y)
  e^{\Lambda(y)}$ is decreasing on $(0,\infty)$.
\end{thm}

\begin{rem}
  Notice that in the case $\beta > 0$ the term in $y^\beta$ inside
  $\Lambda$ does not play any role in the behavior of $g$ as $y \to
  0$,
\end{rem}

We do not prove the asymptotic behavior of self-similar profiles at
$y=0$ in the case $\alpha = 0$ because this was already proved in
\cite{citeulike:1271620}. However, the same techniques employed in
this paper, in particular the rewriting of the equation in section
\ref{sec:self-similar_rewrite}, may be used to give a proof of it
which is somewhat different from the one in
\cite{citeulike:1271620}. Lemma~\ref{lem:II-ss-rewrite} below easily
implies the following additional information on a profile $g$:

\begin{thm}
  \label{thm:II-asymptotic-behavior-intro}
  Assume that the coagulation kernel is of the form
  \eqref{eq:kernel-expression-precise} with $\alpha = 0$. If $g$ is a
  self-similar profile for Smoluchowski's coagulation equation, then
  the function $y \mapsto y^{\tau-1} \int_y^\infty g(z) \,dz$ is
  decreasing on $(0,\infty)$, where
  \begin{equation*}
    \tau := 2 - (1-\lambda) M_\lambda[g].
  \end{equation*}
\end{thm}

We also prove the following partial uniqueness results:
\begin{thm}
  \label{thm:uniqueness-intro-II}
  Consider a coagulation kernel of the form
  \eqref{eq:kernel-expression-precise} with $\alpha = 0$. Assume that
  $g_1$ and $g_2$ are two self-similar profiles \emph{of the same
    mass} for Smoluchowski's equation, and also that $M_\lambda[g_1] =
  M_\lambda[g_2]$. Then, $g_1 = g_2$.
\end{thm}
Here and in the rest of the paper we use the notation $M_\mu[g]$ to
denote the $\mu$-moment of a function $g$, for any $\mu \in \RR$:
\begin{equation}
  \label{eq:moments-notation}
  M_\mu[g] := \int_0^\infty y^\mu g(y) \,dy.
\end{equation}

For coefficients with $\alpha < 0$ our result is:
\begin{thm}
  \label{thm:uniqueness-intro-III}
  Consider a coagulation kernel of the form
  \eqref{eq:kernel-expression-precise} with $\alpha < 0$. Assume that
  $g_1$ and $g_2$ are two self-similar profiles for Smoluchowski's
  equation such that
  \begin{gather}
    M_\alpha[g_1] = M_\alpha[g_2] =: M_\alpha,
    \\
    M_\beta[g_1] = M_\beta[g_2] =: M_\beta,
    \\
    \label{eq:III-limits-equal}
    \lim_{y \to 0} g_1(y) e^{\Lambda(y)}
    =
    \lim_{y \to 0} g_2(y)  e^{\Lambda(y)},
  \end{gather}
  where
  \begin{equation*}
    \Lambda(y) :=
    2 \log y
    - (1-\lambda) \left(
      \frac{M_\beta}{\alpha} y^\alpha
      +
      \frac{M_\alpha}{\beta} y^\beta
    \right).
  \end{equation*}
  Then, $g_1 = g_2$.
\end{thm}

\begin{rem}
  The limit at $y=0$ which appears in this result is proved to exist
  in section \ref{sec:asymptotic-behavior-III}; see
  theorem~\ref{thm:III-asymptotic-behavior-intro} below. Using it it
  is easy to see that condition \eqref{eq:III-limits-equal} in the
  above theorem is equivalent to the requirement that $g_1 \sim g_2$
  when $y \to 0$, this is, $\lim_{y \to 0} g_1(y)/g_2(y) = 1$.
\end{rem}

To our knowledge, no uniqueness result at all was available for
self-similar profiles of Smoluchowski's equation. The natural
conjecture is that the full uniqueness result should hold, this is,
that two scaling profiles with the same mass are necessarily equal.
This does hold in the better understood cases with $a(x,y) = 1$ and
$a(x,y)=x+y$, but the techniques we use here do not seem enough to
show this for general $\lambda$. A central difficulty is the fact that
the moments of the equation ($M_\lambda$ when $\alpha = 0$, or
$M_\alpha, M_\beta$ when $\alpha < 0$) are a global property of the
solution that cannot be computed \emph{a priori} (unlike the $a(x,y) =
1$ or $a(x,y)=x+y$ cases), and which are even not easy to approximate
numerically \cite{citeulike:2298278}.

\subsection{Tools and method of proof}
\label{sec:tools-intro}

Among the tools used to prove the above results we highlight a new
representation of the coagulation operator, which is given in section
\ref{sec:representation}, and a rewriting of the self-similar profile
equation, given in section \ref{sec:self-similar_rewrite}. Let us
comment on them briefly.

In order to rewrite the coagulation operator, define a distribution
associated to any function $f$ of finite mass. We use the Banach space
$L^1_1$ of real measurable functions on $(0,\infty)$ with finite first
moment:
\begin{equation}
  \label{eq:def-L1_1}
  L^1_1 := L^1 ((0,\infty); y \,dy).
\end{equation}
In general, we use the notation
\begin{equation}
  \label{eq:def-L1_k}
  L^1_k := L^1 ((0,\infty); y^k \,dy)
\end{equation}
for $k \in \RR$.

\begin{dfn}[Distribution associated to $f \in L^1_1$]
  \label{dfn:f-distribution}
  Given a function $f \in L^1_1$,
  we define the distribution $\{f\}$
  on $\RR$ as
  \begin{equation}
    \label{eq:def-f-distribution}
    \ap{ \{f\}, \phi}
    :=
    \int_0^\infty f(z) \left( \phi(z) - \phi(0) \right)\,dz
    \quad
    \text{ for } \phi \in \mathcal{C}^\infty_0(\RR).
  \end{equation}
\end{dfn}
Note that when $f$ is not absolutely integrable at $0$, this is just
the classical definition of the finite part of the integral
$\int_0^\infty \phi(y) f(y) \,dy$ \cite{citeulike:1915345}. Here, we
keep the same expression even when $f$ is integrable. Then, for a
coagulation kernel of the form (\ref{eq:kernel-expression-precise}),
the coagulation operator may be written as
\begin{equation*}
  C(g,g) = \{y^\alpha g\}*\{y^\beta g\}.
\end{equation*}
Here we are considering $C(g,g)$ defined as a distribution; for a more
precise statement, see section \ref{sec:representation}.

The above expression has the advantage of being simple and lending
itself to convenient and perfectly rigorous manipulations. For
example, it is easy to recover the known expression for the primitive
of $C(g,g)$, as the convolution above commutes with integration and
derivation operators. Furthermore, it gives some insight into the way
$C(g,g)$ works: if we are working, say, in the case $\alpha = 0$,
where $g(y) \sim K\, y^{-\tau}$ when $y \to 0$ for some $K >0$, one
can see that $C(g,g)$ is in many respects analogous to a fractional
derivative of $g$ of order $\tau - 1$, as a fractional derivative of
this order is precisely the convolution with $\{y^{-\tau}\}$, times a
constant. This is crucially used in the proof of the differentiability
of profiles in section \ref{sec:local-regularity}, where we carry out
a bootstrap argument which shows that if a profile $g$ is $k$ times
differentiable, then the self-similar profile equation implies that it
must be in fact $k+1-\lambda$ times differentiable. The use of
fractional derivatives comes naturally in this context, and actually a
different version of this bootstrap argument has already been used in
\cite{citeulike:1271620} to show that profiles for $\alpha = 0$ are
$\mathcal{C}^1$. There, it was necessary to first show a certain
Hölder regularity of the solution, and then use this information to
obtain $\mathcal{C}^1$ regularity. Formulating this in terms of gain
of fractional derivatives makes it easy to iterate the argument to
obtain $\mathcal{C}^\infty$ regularity and extend it to other kernels.

For the study of the behavior of solutions near $y=0$ and the proof of
uniqueness we rewrite the self-similar profile equation by solving the
differential part, along with any other term which can be separated
and solved; see section \ref{sec:self-similar_rewrite} for a statement
of this.

Finally, let us sketch the idea of the proof of our uniqueness result.
A fundamental obstacle that makes equation \eqref{eq:ss-profile-eq} difficult to
study is the fact that it involves the nonlocal term $C(g,g)$, which
makes it very different from an ordinary differential equation.
However, the coagulation operator for a constant coefficient has a
gain part which only uses values of the function $g$ less than $y$,
and a loss part which is nonlocal only through the appearance of the
integral of $g$. The idea is then to solve the latter part of the
equation, which is simpler, \emph{assuming the value of the involved
  moments is given}, and then look at the remaining part as an
equation which is local \emph{near 0}, to which the kind of arguments
used in the theory of o.d.e.s can be adapted.

This idea works well for coefficients with $\alpha < 0$, as then the
solution decays rapidly near $y=0$, but it is not directly applicable
for coefficients with $\alpha = 0$, as then solutions are known not to
be integrable near $0$ and the gain and loss parts cannot be separated
in the same way (one cannot separate the integral of $g$, as this term
does not make sense). But, one can still find a way to separate the
equation \emph{for the primitive of $g$} in a similar way, and then
carry out the argument on it. This line of reasoning is followed in
section \ref{sec:uniqueness}, where
theorems~\ref{thm:uniqueness-intro-II} and
\ref{thm:uniqueness-intro-III} are proved. This idea depends crucially
on the fact that in the case $\alpha \leq 0$ the self-similar solution
behaves \emph{better} than $y^{-1-\lambda}$ near $y=0$, so that the
linear operator $L_g(f) := C(f,g)$ is regularizing near $y=0$. Hence,
this method does not give new information in the case $\alpha > 0$,
where solutions are expected to be asymptotic to $y^{-1-\lambda}$
\cite{citeulike:1300131}.

\subsection{Organization of the paper}

In the next section we give the basic definitions of self-similar
profile, and precisely define the coagulation operator as a
distribution on $(0,\infty)$. In section \ref{sec:representation} we
give the representation of this operator in the way mentioned above,
which in particular extends it naturally to a distribution on
$(0,\infty)$. This result is a central idea in the proof of infinite
differentiability of scaling profiles, to which section
\ref{sec:local-regularity} is devoted. Section
\ref{sec:self-similar_rewrite} proves that the self-similar profile
equation can be rewritten, as we briefly explained before, in a way
which is very useful to prove the asymptotic behavior of $\alpha < 0$
profiles at $y=0$ (given in
section~\ref{sec:asymptotic-behavior-III}). Finally, we prove our
uniqueness theorems~\ref{thm:uniqueness-intro-II} and
\ref{thm:uniqueness-intro-III} in section~\ref{sec:uniqueness}.

In an appendix (section \ref{sec:appendix}) we include a very brief
introduction to fractional derivatives in order to clarify the
notation and definitions we are using, as they are not completely
standard in the literature. We also prove some simple but delicate
results on fractional differentiation which are used in this paper,
and for which we could not find a reference which gives the explicit
statement.

\section{Preliminaries: self-similar profiles}
\label{sec:preliminaries}

When one wants to define precisely the concept of self-similar
profile, it is a well-known inconvenience that in order for $C(g,g)$
to be well defined by expression (\ref{eq:coag_operator}), $g$ must
meet certain conditions; the ones which are usually imposed are
finiteness conditions on certain moments near $0$ and $\infty$, which
\emph{are not satisfied} by the natural solutions of
(\ref{eq:ss-profile-eq}), known to have a nonintegrable singularity at
$y = 0$ \cite{citeulike:1271620, citeulike:1300131, citeulike:1069567,
  citeulike:1172261}. Hence, the definition of $C(g,g)$ is often
changed for a less restrictive weak formulation by integrating against
a suitably regular test function $\phi$ with compact support on
$(0,\infty)$, which we will do next. Also, $C(g,g)$ is quadratic in
$g$, and we will need to consider its associated symmetric bilinear
operator, so we actually give a weak definition for the latter:
\begin{dfn}
  \label{dfn:coag-operator-usual}
  Let $a$ be a symmetric nonnegative measurable function defined on
  $(0,\infty) \times (0,\infty)$, and $f,g$ be locally integrable
  functions defined on $(0,\infty)$ such that
  \begin{equation}
    \label{eq:condition-def-C-distribution}
    \int_0^\infty \!\!\! \int_0^\infty
    a(y,z) \,y\, z \abs{ f(y) g(z) }\,dy \,dz
    < \infty.
  \end{equation}
  We define \emph{the coagulation operator $C(f,g)$ associated to the
    coagulation coefficient $a$} as the following distribution on
  $(0,\infty)$:
  \begin{multline}
    \label{eq:C-weak-usual}
    \ap{C(f,g), \phi}
    \\
    = \frac{1}{2} \int_0^\infty \!\!\! \int_0^\infty
    a(y,z) f(y) g(z) (\phi(y+z) - \phi(y) - \phi(z)) \,dx \,dz
    \\
    \text{ for } \phi \in \mathcal{C}^\infty_0(0,\infty).
  \end{multline}
\end{dfn}
\begin{rem}
  For $\phi \in \mathcal{C}_0^\infty(0,\infty)$ there is always a
  constant $M > 0$ (which depends on the compact support of $\phi$)
  such that $$\abs{\phi(y+z) - \phi(y) - \phi(z)} \leq C \,y\,z, \quad
  \text{ for } y,z > 0,$$ which can be readily seen by the Mean Value
  Theorem. Then, the integral \eqref{eq:C-weak-usual} is finite by
  \eqref{eq:condition-def-C-distribution}.
\end{rem}
When $f$ is regular enough, $\ap{ C(f,f), \phi}$ can be see to be
equal to $\int_0^\infty \phi(y) C(f,f)(y) \,dy$, where $C(f,f)(y)$ is
given by expression \eqref{eq:coag_operator}. Condition
(\ref{eq:condition-def-C-distribution}) can be somewhat weakened in
some cases by loosening the integrability condition on $f$ at
$+\infty$, but the above one is simpler and will be enough for our
purposes.

\begin{dfn}[Self-similar profile]
  \label{dfn:self-similar-profile}
  Assume that the coagulation coefficient $a$ is homogeneous of degree
  $\lambda$. A nonnegative locally integrable function $g:(0,\infty)
  \to [0,\infty)$ for which
  \begin{gather*}
    \int_0^\infty y\, g(y) \,dy < \infty,
    \\
    \int_0^\infty \!\!\! \int_0^\infty
    a(y,z) \,y\, z \abs{ g(y) g(z) }\,dy \,dz
    < \infty
  \end{gather*}
  is a \emph{self-similar profile} for Smoluchowski's coagulation
  equation if equation (\ref{eq:ss-profile-eq}) holds in the sense of
  distributions on $(0,\infty)$; this is, if
  \begin{equation}
    \label{eq:ss-profile-eq-precise}
    2g + y \,\partial_y g
    + (1-\lambda) C(g,g)
    =
    0.
  \end{equation}
\end{dfn}

\section{A representation of the coagulation operator}
\label{sec:representation}

In this section we want to give a representation of the coagulation
operator when $a$ has the form \eqref{eq:kernel-expression-precise}
which sheds some light on its structure, and in particular will be
very helpful to prove our regularity results later. To begin with, let
us give a natural extension of $C(f,g)$ from definition
\ref{dfn:coag-operator-usual} to a distribution on $\RR$:

\begin{dfn}
  \label{dfn:coag-operator-extension}
  Take a coagulation coefficient $a$ and functions $f,g$ in the
  conditions of definition \ref{dfn:coag-operator-usual}. We define
  the coagulation operator associated to the coagulation coefficient
  $a$, applied to $f,g$, as the distribution $C(f,g)$ on $\RR$ given
  by
  \begin{multline}
    \label{eq:fundamental_identity}
    \ap{ C(f,g), \phi }
    \\
    := \frac{1}{2} \int_0^\infty \int_0^\infty
    a(x,z) f(x) g(z)
    (\phi(x+z) - \phi(x) - \phi(z) + \phi(0)) \,dx \,dz
    \\
    \text{ for all } \phi \in \mathcal{C}^\infty_0(\RR).
  \end{multline}
\end{dfn}
It is easy to see that this is well defined as a distribution on
$\RR$, even with the weak requirements on $f$ and $g$, and that it
extends definition \ref{dfn:coag-operator-usual}. Note the addition of
the term $\phi(0)$, which does not make a difference when $\phi$ has
compact support on $(0,\infty)$; later we will see how this extension
comes about naturally. To give our representation for $C(f,g)$ we will
use the notation from definition \ref{dfn:f-distribution}. Let us
initially treat the case of a constant coefficient $a \equiv 1$, and
then see how it extends to other coefficients:

\begin{prp}
  \label{thm:representation-C0}
  Take $f,g \in L^1_1$, and let $C_0$ be the coagulation operator with
  a constant coefficient $a \equiv 1$ as given in definition
  \ref{dfn:coag-operator-extension}. Then,
  \begin{equation*}
    C_0(f,g) = \frac{1}{2} \{f\} * \{g\}
    \quad
    \text{ as distributions on } \RR.
  \end{equation*}
\end{prp}
Here, the convolution $\{f\}*\{g\}$ is understood as a convolution of
distributions with compact support to the left
\cite{citeulike:1915345}.  This expression is surprisingly simple, and
is well-suited for the study of the coagulation operator when the
functions $f$, $g$ have a singularity at $0$, which is the case with
some self-similar profiles. Its proof just consists of writing out the
definitions:

\begin{proof}
  For $\phi \in \mathcal{C}^\infty_0(\RR)$,
  \begin{equation}
    \label{eq:e1}
    \ap{\{f\}*\{g\}, \phi}
    =
    \ap{\{f\}, (\mathcal{R}\{g\}) * \phi},
  \end{equation}
  (where $\mathcal{R}$ is the reflection operator, $\mathcal{R} \phi
  (y) := \phi(-y)$, defined by duality on distributions) and
  \begin{equation*}
    ((\mathcal{R}\{g\}) * \phi) (x)
    =
    \ap{\{g\}, \tau_{-x} \phi}
    =
    \int_{0}^\infty g(z) (\phi(x+z) - \phi(x)) \,dz,
  \end{equation*}
  so from \eqref{eq:e1} we have
  \begin{multline*}
    \ap{\{f\}*\{g\}, \phi}
    =
    \int_0^\infty f(x)
    \big(
      ((\mathcal{R}\{g\}) * \phi) (x) - ((\mathcal{R}\{g\}) * \phi) (0)
    \big) \,dx
    \\
    =
    \int_0^\infty f(x)
    \Bigg(
      \int_{0}^\infty g(z) (\phi(x+z) - \phi(x)) \,dz
    \\
      -
      \int_{0}^\infty g(z) (\phi(z) - \phi(0)) \,dz
    \Bigg)
    \,dx
    \\
    =
    \int_0^\infty \int_0^\infty
    f(x) g(z)
    \left(
      \phi(x+z) - \phi(x) - \phi(z) + \phi(0)
    \right)
    \,dx \,dz,
  \end{multline*}
  which is our result, in view of expression
  \eqref{eq:fundamental_identity}.
\end{proof}

This directly gives a representation of the coagulation operator
$C(f,g)$ with a coagulation kernel $a$ satisfying
\eqref{eq:kernel-expression-precise}, as the following relation holds
for any $f,g$ satisfying \eqref{eq:condition-def-C-distribution}:
\begin{equation*}
  C(f,g) = C_0(y^\alpha f, y^\beta g) + C_0(y^\beta f, y^\alpha g),
\end{equation*}
where $C(f,g)$ is the operator associated to $a$, and the equality is
an equality of distributions on $\RR$.

\begin{thm}
  \label{thm:representation-C}
  Assume that the coagulation operator $a$ is of the form
  (\ref{eq:kernel-expression-precise}), and take $f, g$ which satisfy
  \eqref{eq:condition-def-C-distribution}. Then, the coagulation
  operator associated to $a$ (as given in definition
  \ref{dfn:coag-operator-extension}) can be written as
  \begin{equation*}
    C(f,g) = \frac{1}{2} \left(
      \{y^\alpha f\}*\{y^\beta g\}
      + \{y^\beta f\}*\{y^\alpha g\}
    \right),
  \end{equation*}
  where equality holds as distributions on $\RR$.
\end{thm}

We emphasize that this operator is defined as a distribution on $\RR$
and, as pointed out after definition
\ref{dfn:coag-operator-extension}, extends the usual operator $C$ from
definition \ref{dfn:coag-operator-usual} (which is a distribution on
$(0,\infty)$).

\section{Local regularity}
\label{sec:local-regularity}


With the representation of the coagulation operator $C(g,g)$ given in
theorem~\ref{thm:representation-C}, the study of its regularity can be
viewed as the study of the regularity of convolutions of the above
type, which is a more manageable problem. The difficulty here is that
when $\alpha \geq 0$ a scaling solution $g$ is not integrable near
$0$, but is expected to be very regular locally, and hence we need to
study the convolution of functions which have singularities at $0$.
Precisely, the following general result on the local integrability of
scaling profiles near $y=0$ is known (see \cite{citeulike:1172261} or
\cite{citeulike:1069567} for a proof):

\begin{prp}
  \label{prp:finite-moments}
  Assume that the coagulation coefficient $a$ is of the form
  \eqref{eq:kernel-expression-precise}, or is a finite linear
  combination of terms of that form, all with the same homogeneity
  degree $\lambda$. Then, all self-similar profiles $g$ for
  Smoluchowski's equation (in the sense of definition
  \ref{dfn:self-similar-profile}) satisfy that
  \begin{equation*}
    \int_0^R y^k g(y) \,dy < \infty
    \quad \text{ for all } R > 0
    \text{ and all } k > \lambda.
  \end{equation*}
\end{prp}

For convenience, we will measure the regularity of a function by
looking at how many of its derivatives are locally integrable on
$(0,\infty)$. To study the regularity of $C$ we will need to use an
interesting relationship between the kind of singularity of a function
near $0$ and the local integrability of its fractional integrals,
which we give in lemma~\ref{lem:D-mu_bound} below. Let us start with
the following elementary lemma that we state without proof:

\begin{lem}
  \label{lem:difference_integral}
  For $0 < k < 1$,
  \begin{equation}
    \int_0^\infty ( x^{k-1} - (z+x)^{k-1})\,dx
    = \frac{1}{k} z^k
    \quad \text{ for } z > 0
    .
  \end{equation}
\end{lem}


\begin{lem}
  \label{lem:D-mu_bound}
  If $f \in L^1_k$, with $0 < k \leq 1$, then
  \begin{equation*}
    \norm{D^{-k}\{f\}}_{L^1(\RR)}
    \leq
    \frac{2}{\Gamma(k+1)}
    \norm{y^k f}_{L^1(0,\infty)}.
  \end{equation*}
\end{lem}

\begin{rem}
  We recall that we denote $L^1_k := L^1((0,\infty); y^k \,dy)$ as in
  eq. \eqref{eq:def-L1_k}.
\end{rem}

\begin{proof}
  For any $\phi \in \mathcal{C}^\infty_0(\RR)$ we will prove that
  \begin{equation}
    \label{eq:in-1}
    \abs{ \ap{ D^{-k} \{f\}, \phi} }
    \leq
    \frac{2}{\Gamma(k+1)}
    \norm{y^k f}_{L^1(0,\infty)} \norm{\phi}_\infty,
  \end{equation}
  which is equivalent to our inequality. We have:
  \begin{equation}
    \label{eq:in-2}
    \ap{ D^{-k} \{f\}, \phi}
    =
    \ap{ \{f\}, D_{-k} \phi}
    =
    \int_0^\infty f(z)
    \left(
      D_{-k} \phi (z) - D_{-k} \phi(0)
    \right) \,dz,
  \end{equation}
  and the part inside the parentheses is
  \begin{multline}
    \label{eq:in-3}
    D_{-k} \phi (z) - D_{-k} \phi(0)
    \\
    =
    \frac{1}{\Gamma(k)}
    \int_z^\infty \phi(x) (x-z)^{k-1} \,dx
    -
    \frac{1}{\Gamma(k)}
    \int_0^\infty \phi(x) x^{k-1} \,dx
    \\
    =
    \frac{1}{\Gamma(k)}
    \int_z^\infty \phi(x)
    \left(
      (x-z)^{k-1} - x^{k-1}
    \right) \,dx
    -
    \frac{1}{\Gamma(k)}
    \int_0^z \phi(x) x^{k-1} \,dx.
  \end{multline}
  We put back this these two terms in \eqref{eq:in-2}, and bound them
  separately. The first term is $0$ when $k=1$, and for $0 < k <1$ we
  have
  \begin{multline}
    \label{eq:in-4}
    \int_0^\infty \abs{f(z)}
    \int_z^\infty \abs{\phi(x)}
    \abs{
      (x-z)^{k-1} - x^{k-1}
    }
    \,dx
    \,dz
    \\
    \leq
    \norm{\phi}_\infty
    \int_0^\infty \abs{f(z)}
    \int_z^\infty
    \abs{
      (x-z)^{k-1} - x^{k-1}
    }
    \,dx
    \,dz
    \\
    =
    \frac{1}{k}
    \norm{\phi}_\infty
    \int_0^\infty z^k \abs{f(z)}
    \,dz,
  \end{multline}
  thanks to lemma~\ref{lem:difference_integral}. As for the second
  term in \eqref{eq:in-3}, putting it into \eqref{eq:in-2} we have
  \begin{multline}
    \label{eq:in-5}
    \int_0^\infty \abs{f(z)}
    \int_0^z \abs{\phi(x)} x^{k-1} \,dx
    \,dz
    \leq
    \norm{\phi}_\infty
    \int_0^\infty \abs{f(z)}
    \int_0^z x^{k-1} \,dx
    \,dz
    \\
    =
    \frac{1}{k}
    \norm{\phi}_\infty
    \int_0^\infty z^k \abs{f(z)}
    \,dz.
  \end{multline}
  Then, eqs. \eqref{eq:in-2}--\eqref{eq:in-5} prove that
  \begin{equation*}
    \abs{ \ap{ D^{-k} \{f\}, \phi} }
    \leq
    \frac{2}{k \Gamma(k) }
    \norm{\phi}_\infty
    \int_0^\infty z^k \abs{f(z)}
    \,dz,
  \end{equation*}
  which proves inequality \eqref{eq:in-1}, taking into account that $k
  \Gamma(k) = \Gamma(k+1)$.
\end{proof}

In the light of the above lemma, our next result can be understood as
saying: if two functions are locally regular but have a nonintegrable
singularity at 0, their convolution is slightly less
regular. \emph{How much} less regular it is depends on the nature of
the singularity. In particular, the local regularity of the
convolution depends only on local properties of the initial functions,
which is a general property of the convolution operation. In the next
lemma, the reader can keep in mind that $\mu$ will be negative when we
use it, so a function $f$ for which $D^{\mu}f$ is integrable may well
be not integrable near $0$, as lemma~\ref{lem:D-mu_bound} makes clear.

\begin{lem}
  \label{lem:regularity-of-convolution}
  Let $T$, $S$ be two distributions on $\RR$ with support on
  $[0,\infty)$ (this is, $T,S \in \mathcal{D}_L'$), and assume that
  \begin{enumerate}
  \item For some $\nu \in \RR$, $D^\nu T$ and $D^\nu S$ are locally
    integrable on $(0,\infty)$.
  \item For some $\mu \leq \nu$, $D^{\mu} T$, $D^\mu S$ are locally
    integrable on $\RR$.
  \end{enumerate}
  Then, the distribution $D^{\mu + \nu}(T * S)$ is locally integrable
  on $(0,\infty)$.
\end{lem}

\begin{proof}
  We break $T$ and $S$ into a part near $0$, an intermediate part, and
  a part near $\infty$. For this, choose $0 < \epsilon < 1/4$. We can
  find smooth nonnegative cutoff functions $\Phi_0$, $\Phi_1$,
  $\Phi_2$ on $(0,\infty)$ such that
  \begin{align*}
    &\Phi_0 \equiv 1 \text{ on } (0, \epsilon),
    \qquad \Phi_0 \equiv 0 \text{ on } (2\epsilon, \infty)
    \\
    &\Phi_1 \equiv 1 \text{ on } (2\epsilon, \frac{1}{\epsilon}),
    \qquad \Phi_1 \equiv 0
    \text{ on } (0,\epsilon) \cup (\frac{2}{\epsilon}, \infty)
    \\
    &\Phi_2 \equiv 1 \text{ on } (\frac{2}{\epsilon}, \infty),
    \qquad \Phi_2 \equiv 0 \text{ on } (0, \frac{1}{\epsilon})
  \end{align*}
  and such that
  \begin{equation*}
    \Phi_0 + \Phi_1 + \Phi_2 \equiv 1
    \text{ on } (0,+\infty).
  \end{equation*}
  In other words, $\Phi_0, \Phi_1, \Phi_2$ form a partition of unity on
  $(0,\infty)$ subordinated to the open cover $(0, 2\epsilon) \cup
  (\epsilon, 2/\epsilon) \cup (1/\epsilon, \infty)$. Then,
  \begin{gather*}
    S
    =
    S \Phi_0 + S \Phi_1 + S \Phi_2
    =: S_0 + S_1 + S_2,
    \\
    T
    =
    T \Phi_0 + T \Phi_1 + T \Phi_2
    =: T_0 + T_1 + T_2,
  \end{gather*}
  where we have denoted $S_i := S \Phi_i$, $T_i := T \Phi_i$ for $i =
  0,1,2$. We can break the convolution $S * T$ by using this
  decomposition. Note that for $i=0,1,2$, $S_2 * T_i$ is zero on
  $(0,1/\epsilon)$, and the same happens with $S$, $T$ interchanged;
  as we are only interested in studying the regularity of $S*T$ on a
  bounded interval, we can disregard these terms and write
  \begin{equation*}
    S * T
    =
    S_0 * T_0
    +
    S_0 * T_1
    +
    S_1 * T_0
    +
    S_1 * T_1
    \quad \text{ on } (0, 1/\epsilon).
  \end{equation*}
  Similarly, the term $S_0 * T_0$ is zero on $(4\epsilon, \infty)$, so
  we have
  \begin{equation}
    \label{eq:local-decomposition}
    S * T
    =
    S_0 * T_1
    +
    S_1 * T_0
    +
    S_1 * T_1
    \quad \text{ on } (4\epsilon, 1/\epsilon).
  \end{equation}
  Then, we can write $D^{\mu + \nu}$ of each of these terms by using
  theorem~\ref{thm:convolution-derivative-distributions}:
  \begin{gather}
    \label{eq:dec-1}
    D^{\mu+\nu}(S_0 * T_1) = (D^\mu S_0) * (D^\nu T_1)
    \\
    \label{eq:dec-2}
    D^{\mu+\nu}(S_1 * T_0) = (D^\nu S_1) * (D^\mu T_0)
    \\
    \label{eq:dec-3}
    D^{\mu+\nu}(S_1 * T_1) = (D^\mu S_1) * (D^\nu T_1).
  \end{gather}
  By the hypotheses of the lemma, we can see that all of the terms
  that take part in the convolutions on the right hand side are
  integrable functions, as the product by $\Phi_0$ or $\Phi_1$ does
  not change their local regularity properties
  (theorem~\ref{thm:local-regularity-of-product}). Let us do the
  reasoning for $D^\mu S_1$: as $D^\nu S$ is integrable on
  $(4\epsilon, 1/\epsilon)$ by hypothesis, we have that $D^\nu (S
  \Phi_1) = D^\nu S_1$ is integrable on that interval by
  theorem~\ref{thm:local-regularity-of-product}; then, thanks to
  lemma~\ref{lem:less-derivatives-better}, $D^\mu S_1$ is also, as
  $\mu \leq \nu$. The rest of the terms can be treated analogously,
  and are seen to be integrable without the help of
  lemma~\ref{lem:less-derivatives-better}.
  
  Then, all the terms on the right hand side of
  (\ref{eq:dec-1})--(\ref{eq:dec-3}) are convolutions of integrable
  functions, and hence are integrable, and
  (\ref{eq:local-decomposition}) then proves that $D^{\mu+\nu}(S*T)$
  is integrable on $(4\epsilon, 1/\epsilon)$.
\end{proof}

\begin{prp}
  \label{prp:regularity_of_C}
  Take a coagulation coefficient $a$ which is of the form
  \eqref{eq:kernel-expression-precise} with $\alpha \geq 0$, and $C$
  the coagulation operator associated to $a$ (given by definition
  \ref{dfn:coag-operator-extension}). Assume that $g: (0,\infty) \to
  \RR$ is such that
  \begin{enumerate}
    \label{item:g-moment}
  \item $y^k g$ is locally integrable on $[0,\infty)$ for some $1 \geq
    k \geq \beta$,
  \item and $D^\nu g$ is locally integrable on $(0,\infty)$ for some
    $\nu \geq 0$.
  \end{enumerate}
  Then, $D^{\alpha - k + \nu} C(g,g)$ is locally integrable on
  $(0,\infty)$.
\end{prp}

\begin{rem}
  The hypothesis that $\alpha \geq 0$ is given for convenience, as we
  only need the result in that case; however, the lemma is true and
  proved in the same way also for negative $\alpha$ with the
  additional requirement that $1+\alpha \geq k \geq \beta$, so that
  $\{y^\alpha g\}$ makes sense according to definition
  \ref{dfn:f-distribution}.
\end{rem}

\begin{rem}
  The result works also when $a$ is a linear combination of terms of
  the form \eqref{eq:kernel-expression-precise} with the same
  $\lambda$. In this case, if $\alpha_i, \beta_i$ are the exponents in
  the $i$-th term in the linear combination, the proposition is true
  taking $\beta := \max_i \{\beta_i\}$ and $\alpha := \min_i
  \{\alpha_i\}$.
\end{rem}

\begin{proof}
  It is enough to prove it for a coagulation coefficient of the form
  (\ref{eq:kernel-expression-precise}) with $\alpha \geq 0$, as then
  we can apply the result to each term of the linear combination. For
  such a coefficient, the representation
  theorem~\ref{thm:representation-C} shows we can write $C(g,g)$ as
  \begin{equation*}
    C(g,g) = \{y^\alpha g\} * \{y^\beta g\}.
  \end{equation*}
  As $y \mapsto y^k g(y)$ is locally integrable on $[0,\infty)$, we
  know that
  \begin{itemize}
  \item $y^{k - \alpha} y^\alpha g$ is locally integrable on
    $[0,\infty)$, and
  \item $y^{k - \beta} y^\beta g$ is locally integrable on
    $[0,\infty)$,
  \end{itemize}
  and hence from lemma~\ref{lem:D-mu_bound}
  \begin{itemize}
  \item $D^{\alpha - k} \{y^\alpha g\}$ is locally integrable on
    $\RR$, and
  \item $D^{\beta - k} \{y^\beta g\}$ is locally integrable on $\RR$,
    so $D^{\alpha - k} \{y^\beta g\}$ is also (as $\beta \geq \alpha$;
    see lemma~\ref{lem:less-derivatives-better}).
  \end{itemize}
  In addition, both $D^\nu \{y^\alpha g\}$ and $D^\nu \{y^\beta g\}$
  are locally integrable on $(0,\infty)$, as $\{y^\alpha g\}$,
  $\{y^\beta g\}$ are equal to the functions $y^\alpha g$, $y^\beta
  g$, respectively, on that set, and then
  theorem~\ref{thm:local-regularity-of-product} applies there. Hence,
  we obtain our result as an application of
  lemma~\ref{lem:regularity-of-convolution} with $S := \{y^\alpha
  g\}$, $T := \{y^\beta g\}$ and $\mu := \alpha - k$
\end{proof}

Now we can finally prove theorem~\ref{thm:local-regularity-intro}:


\begin{proof}[Proof of theorem~\ref{thm:local-regularity-intro}]
  Take any $\lambda < k < 1$, with $\lambda$ the homogeneity degree of
  $a$. Then, $y^k g$ is locally integrable on $[0,\infty)$ as recalled
  in proposition~\ref{prp:finite-moments} (actually, we know it is
  integrable). We will show the following: if, for some $\nu \geq 0$,
  $D^\nu g$ is locally integrable on $(0,\infty)$, then $D^{\nu + 1 +
    \alpha - k} g$ is also locally integrable on $(0,\infty)$. As $\nu
  + 1 + \alpha - k > \nu$, this implies that $g$ is infinitely
  differentiable by a bootstrap argument starting with $\nu = 0$ ($g$
  is locally integrable by definition).

  To show this, write the equation for a self-similar profile as
  \begin{equation}
    \label{eq:scaling-eq}
    2 g(y) + y D^1 g + (1-\lambda) C(g,g) = 0
    \quad \text{ as distributions on } (0,\infty).
  \end{equation}
  Assume that $D^\nu g$ is locally integrable on $(0,\infty)$. Then,
  by proposition~\ref{prp:regularity_of_C}, $D^{\nu + \alpha - k}
  C(g,g)$ is locally integrable on $(0,\infty)$, and hence eq.
  (\ref{eq:scaling-eq}) shows that $D^{\nu + \alpha - k} (y D^1 g)$ is
  locally integrable on $(0,\infty)$. By
  theorem~\ref{thm:local-regularity-of-product}, the same is true of
  $D^{\nu + \alpha - k} D^1 g = D^{\nu + \alpha - k +1} g$, which
  proves our claim. Hence, $g$ is infinitely differentiable.
\end{proof}

\section{Rewriting the self-similar equation}
\label{sec:self-similar_rewrite}

One of the techniques that we use in order to study the behavior of
scaling profiles near $y=0$ and the uniqueness of self-similar
solutions is a way of rewriting equation
\eqref{eq:ss-profile-eq-precise} in which we ``solve'' the
differential part of the equation as far as possible. We introduce
this method next:

\begin{lem}[Solution of an o.d.e.]
  \label{lem:ode-solution}
  Let $g$ be an absolutely continuous function, and let $h$, $\mu$ be
  locally integrable functions, all of them defined on
  $(0,\infty)$. If the following equation holds
  \begin{equation}
    \label{eq:ode-g}
    \mu(y) g(y) + y g'(y) = h(y)
    \quad \text{ for almost all } y > 0,
  \end{equation}
  then $g$ is given by
  \begin{equation*}
    g(y) = K(y) e^{-\Lambda(y)}
    \quad \text{ for all } y > 0,
  \end{equation*}
  where $\Lambda$ and $K$ are absolutely continuous functions which
  satisfy that
  \begin{gather}
    \label{eq:derivative_Lambda}
    \Lambda'(y) = \frac{\mu(y)}{y}
    \quad
    \text{ for almost all } y > 0,
    \\
    \label{eq:derivative_K}
    K'(y) = \frac{1}{y} e^{\Lambda(y)} h(y)
    \quad
    \text{ for almost all } y > 0.
  \end{gather}
\end{lem}

\begin{proof}
  We remark that one may find the expression of $g$ by the method of
  variation of constants. To prove the result, a direct check shows
  that if we take functions $\Lambda$, $K$ satisfying
  (\ref{eq:derivative_Lambda}), (\ref{eq:derivative_K}) and define
  \begin{equation*}
    \tilde{g}(y) := K(y) e^{-\Lambda(y)},
  \end{equation*}
  then equation (\ref{eq:ode-g}) holds with $\tilde{g}$ instead of
  $g$. We may add a constant to $K$ so that $g(1) = \tilde{g}(1)$, for
  instance. Now, if we regard (\ref{eq:ode-g}) as an ordinary
  differential equation for $g$, then both $g$ and $\tilde{g}$ are
  solutions of it in the sense of Carathéodory, and then general
  uniqueness theorems (see, e.g., \cite{citeulike:2364173}) prove that
  $g = \tilde{g}$.
\end{proof}

\begin{lem}[Primitive of $C$]
  \label{lem:primitive_C}
  Assume that the coagulation coefficient $a$ is of the form
  \eqref{eq:kernel-expression-precise} with $\alpha = 0$, and take a
  function $g \in L^1_1 \cap L^1_\lambda$ (this is, with finite mass
  and finite moment of order $\lambda$). Then the primitive of
  $C(g,g)$ can be written as
  \begin{equation}
    \label{eq:primitive_C}
    D_{-1} C(g,g)
    = G * (y^\lambda g) -  M_\lambda[g] G,
  \end{equation}
  where $G$ is the function given by
  \begin{equation}
    \label{eq:def-G}
    G(y) := \int_y^\infty g(z) \,dz
    \quad \text{ for } y > 0,
    \qquad
    G(y) := 0 \quad \text{ for } y \leq 0.
  \end{equation}
  We remark that this equality is an equality of distributions on
  $\RR$.
\end{lem}

\begin{rem}
  This is not a new result: one may arrive at the same expression by
  taking a characteristic function of the interval $(y,\infty)$ as the
  function $\phi$ in \eqref{eq:C-weak-usual}. The main point of the
  lemma is that it is rigorously proved as an equality of
  distributions.
\end{rem}

\begin{proof}
  The representation of $C(g,g)$ in theorem~\ref{thm:representation-C}
  gives
  \begin{equation}
    C(g,g)
    = \{g\} * \{y^\lambda g\}
    = \{g\} * (y^\lambda g) - M_\lambda[g] \{g\},
  \end{equation}
  as $g$ has finite $\lambda$-moment. Hence, taking the primitive
  $D_{-1}$ (see eq. \eqref{eq:def_derivative_distribution-right}) we
  have
  \begin{multline*}
    D_{-1} C(g,g)
    = (D_{-1}\{g\}) * (y^\lambda g) - M_\lambda[g]  D_{-1} \{g\}
    \\
    = G * (y^\lambda g) - M_\lambda[g] G,
  \end{multline*}
  where we have used the derivation rule for a convolution (cf.
  theorem~\ref{thm:convolution-derivative-distributions}) and the fact
  that, as distributions on $\RR$,
  \begin{equation*}
    D_{-1} \{g\} = G.
  \end{equation*}
\end{proof}

\begin{lem}
  \label{lem:II-ss-rewrite}
  Assume that the coagulation coefficient $a$ is of the form
  \eqref{eq:kernel-expression-precise} with $\alpha = 0$. If $g$ is
  a self-similar solution of Smoluchowski's equation, it holds that
  \begin{equation}
    \label{eq:II-ss-rewrite}
    G(y) = y^{1-\tau} K(y),
  \end{equation}
  for some absolutely continuous function $K$, where $G$ is given by
  \eqref{eq:def-G}, and where
  \begin{gather}
    \label{eq:def-tau}
    \tau := 2 - (1-\lambda) M_\lambda[g]
    \\
    \label{eq:dK-in-lemma}
    K'(y) = - y^{\tau-2} h(y)
    \quad \text{ for almost all } y > 0,
    \\
    \label{eq:def-h}
    h := (1-\lambda)
    \left(
      G * (y^\lambda g)
    \right),
  \end{gather}
  We note that for the convolution in (\ref{eq:def-h}) it is assumed
  that $g(y)$ is $0$ for $y < 0$.
\end{lem}

\begin{rem}
  Note that the quantity $\tau$ here is in agreement with that in
  \cite{citeulike:1300131,citeulike:1271620}.
\end{rem}

\begin{proof}
  It is known that in the case $\alpha = 0$ the moment of order
  $\lambda$ of a solution $g$ is finite
  \cite{citeulike:1069567,citeulike:1271620}. Hence, using
  \eqref{eq:primitive_C} we can rewrite equation
  \eqref{eq:ss-profile-eq} by taking its primitive:
  \begin{equation}
    \label{eq:ss-primitive-1}
    G - y g +
    (1-\lambda)
    \left(
      G * (y^\lambda g) -  M_\lambda[g] G
    \right)
    = 0,
  \end{equation}
  which holds for all $y > 0$. Equivalently,
  \begin{equation}
    \label{eq:ss-primitive-2}
    \left(
      1- (1-\lambda) M_\lambda[g]
    \right) G - y g +
    (1-\lambda) \left(
      G * (y^\lambda g)
    \right)
    = 0.
  \end{equation}
  Rewrite this as
  \begin{equation}
    \label{eq:ss-primitive-3}
    (\tau - 1) G - y g + h
    = 0,
  \end{equation}
  where $\tau$ and $h$ are given by (\ref{eq:def-tau}) and
  (\ref{eq:def-h}).
  Now, if we solve for $G$ in eq. \eqref{eq:ss-primitive-3} by using
  lemma~\ref{lem:ode-solution} with the independent term $h$, we
  obtain \eqref{eq:II-ss-rewrite} and \eqref{eq:dK-in-lemma} in our
  result.
\end{proof}

\begin{lem}
  \label{lem:III-ss-rewrite}
  Assume that the coagulation coefficient $a$ is of the form
  \eqref{eq:kernel-expression-precise} with $\alpha < 0$, and take a
  self-similar solution $g$ of Smoluchowski's equation. Then it holds
  that
  \begin{equation}
    \label{eq:III-ss-rewrite}
    g(y) = K(y) e^{-\Lambda(y)}
  \end{equation}
  for some absolutely continuous function $K$ such that
  \begin{equation}
    \label{eq:III-dmu}
    K'(y) = \frac{1}{y} e^{\Lambda(y)} h(y)
    \quad
    \text{ for almost all } y > 0,
  \end{equation}
  with
  \begin{gather}
    \label{eq:III-def-h}
    h :=
    -(1-\lambda) (y^\alpha g) * (y^\beta g),
    \\
    \label{eq:III-Lambda}
    \Lambda(y) :=
    2 \log y
    - (1-\lambda) \left(
      \frac{M_\beta}{\alpha} y^\alpha
      +
      \frac{M_\alpha}{\beta} y^\beta
      \right)
    \end{gather}
    when $\beta > 0$, and
    \begin{equation}
      \label{eq:III-Lambda-beta=0}
      \Lambda(y) :=
      (2 - (1-\lambda)M_\alpha) \log y
      - (1-\lambda) \frac{M_\beta}{\alpha} y^\alpha
    \end{equation}
    when $\beta = 0$.
\end{lem}


\begin{proof}
  Let $g$ be a self-similar profile for Smoluchowski's coagulation
  equation with such a kernel $a$. It is known
  \cite{citeulike:1069567} that it is infinitely differentiable and
  has finite moments of all orders, and hence satisfies equation
  (\ref{eq:ss-profile-eq}) in a strong way:
  \begin{equation*}
    2g + y \partial_y g
    + (1-\lambda) C(g,g)
    =
    0,
  \end{equation*}
  or, separating the gain and loss parts of $C(g,g)$,
  \begin{multline}
    \label{eq:III-2}
    2g + y \partial_y g
    + (1-\lambda) (y^\alpha g) * (y^\beta g)
    \\
    - (1-\lambda) M_\beta\, y^\alpha g
    - (1-\lambda) M_\alpha\, y^\beta g
    =
    0,
  \end{multline}
  where $M_\alpha$ and $M_\beta$ are the moments of order $\alpha$ and
  $\beta$, respectively, of $g$. Now, if we apply lemma~\ref{lem:ode-solution} to equation
  (\ref{eq:III-2}) with the independent term $h$ given by (\ref{eq:III-def-h})
  and $\mu := 2 - (1-\lambda) ( M_\beta\, y^\alpha + M_\alpha\,
  y^\beta)$, we obtain that
  \begin{equation}
    \label{eq:III-rewrite-proof}
    g(y) = K(y) e^{-\Lambda(y)},
  \end{equation}
  for some absolutely continuous functions $K$, $\Lambda$ such that
  \begin{gather}
    \label{eq:III-dLambda}
    \Lambda'(y) = \frac{1}{y}
    \left(
      2 - (1-\lambda) (M_\beta\, y^\alpha + M_\alpha\, y^\beta )
    \right)
    \quad
    \text{ a.e. } y > 0,
    \\
    K'(y) = \frac{1}{y} e^{\Lambda(y)} h(y)
    \quad
    \text{ a.e. } y > 0.
  \end{gather}
  We may actually choose $\Lambda$ as a particular primitive, as then
  the integration constant for $K$ can be adjusted so that
  (\ref{eq:III-rewrite-proof}) is still true. Then, we can take
  $\Lambda$ as in (\ref{eq:III-Lambda}) (or
  \eqref{eq:III-Lambda-beta=0} when $\beta=0$), and the result is
  proved.
\end{proof}

\section{Asymptotic behavior at $y=0$ for kernels with $\alpha < 0$}
\label{sec:asymptotic-behavior-III}

In this section we prove theorem
\ref{thm:III-asymptotic-behavior-intro}. Assume that the coagulation
coefficient $a$ is of the form \eqref{eq:kernel-expression-precise}
with $\alpha < 0$. Let $g$ be a self-similar profile for
Smoluchowski's coagulation equation with such a kernel $a$. Then from
lemma~\ref{lem:III-ss-rewrite} we know it holds that
\begin{equation}
  \label{eq:III-10}
  g(y) = K(y) e^{-\Lambda(y)},
\end{equation}
for some absolutely continuous functions $K$ such that
\eqref{eq:III-dmu}--\eqref{eq:III-def-h} hold, and with $\Lambda$
given by \eqref{eq:III-Lambda} (or \eqref{eq:III-Lambda-beta=0} when
$\beta = 0$).


We will prove that $K$ is bounded on $(0,R)$ for any $R >
0$. Obviously, $K$ is bounded on any interval $(\delta, R)$ with $0 <
\delta < R$, as is clear from (\ref{eq:III-10}), so the point is in
proving that $K$ is bounded on $(0,\delta)$ for some $\delta > 0$.

Then, take $\epsilon, \delta > 0$ and define
\begin{equation*}
  N_\epsilon := \sup_{y \in (0,\delta)} g(y) \Phi_\epsilon(y),
\end{equation*}
where
\begin{equation}
  \label{eq:III-rig-phi_epsilon}
  \Phi_\epsilon(y) :=
  \begin{cases}
    e^{\Lambda(\epsilon)} & \text{ if } 0 < y < \epsilon
    \\
    e^{\Lambda(y)}  & \text{ if } \epsilon \leq y.
  \end{cases}
\end{equation}
The dependence of $N_\epsilon$ on $\epsilon$ is explicitly noted
because we want to take the limit $\epsilon \to 0$; of course, $N_\epsilon$
depends also on $\delta$, but we do not write the dependence
explicitly, as our intention is to fix $\delta$ at some value, which
has not been chosen yet. We can give a bound for $g$ in terms of
$N_\epsilon$:
\begin{equation}
  \label{eq:rig-g-bound}
  g(y)
  \leq
  N_\epsilon \Phi_\epsilon(y)^{-1}
  \quad \text{ for } y \in (0,\delta).
\end{equation}

On the other hand, from \eqref{eq:III-dmu},
\begin{equation}
  \label{III-rig-0}
  K(y) =
  K(\delta)
  - (1-\lambda)
  \int_y^\delta \frac{1}{z} e^{\Lambda(z)} h(z) \,dz
  \quad \text{ for } y > 0.
\end{equation}
Let us find a bound for $h$ using \eqref{eq:rig-g-bound} and our
knowledge that $g$ is a bounded function \cite{citeulike:1069567}:
\begin{equation}
  \label{eq:III-g-bounded-absolute}
  g(y) \leq K_0
  \quad \text{ for } y > 0.
\end{equation}
We have
\begin{equation}
  \label{eq:III-rig-h}
  h =  -(1-\lambda) (y^\alpha g) * (y^\beta g).
\end{equation}
To bound this for $y \in (0,\delta)$, take $\delta$ small enough so
that $e^{-\Lambda(y)}$ is increasing on $(0,\delta)$ (and so is
$\Phi_\epsilon(y)^{-1}$), and then
\begin{multline}
  \label{eq:III-h-1}
  (y^\alpha g) * (y^\beta g)
  \leq
  N_\epsilon K_0 \,\Phi_\epsilon(y)^{-1}\,  (y^\alpha ) * (y^\beta)
  =
  N_\epsilon K_1 \, \Phi_\epsilon(y)^{-1} y^{\lambda + 1} ,
\end{multline}
for $y \in (0,\delta)$. Then, from \eqref{eq:III-rig-h},
\begin{equation}
  \label{eq:III-rig-h-2}
  \abs{h(y)}
  \leq
  N_\epsilon K_2 \, \Phi_\epsilon(y)^{-1} y^{\lambda + 1},
\end{equation}
and continuing from \eqref{III-rig-0}, taking into account that $e^{\Lambda(z)}
  \Phi_\epsilon(z)^{-1}$ is decreasing on $(0,\delta)$,
\begin{multline}
  \label{eq:III-rig-3}
  \abs{K(y)}
  \leq
  K(\delta)
  + (1-\lambda)
  N_\epsilon K_2 \int_y^\delta z^{\lambda} e^{\Lambda(z)}
  \Phi_\epsilon(z)^{-1} \,dz
  \\
  \leq
  K(\delta)
  +
  N_\epsilon K_3\,
  e^{\Lambda(y)}\,
  \Phi_\epsilon(y)^{-1}\,
  \int_y^\delta z^{\lambda}  \,dz
  \\
  \leq
  K(\delta)
  + N_\epsilon K_4\,
  e^{\Lambda(y)}\,
  \Phi_\epsilon(y)^{-1}\,
  \delta^{\lambda+1}
  \quad \text{ for } y \in (0,\delta).
\end{multline}
Hence, multiplying by $e^{-\Lambda(y)} \Phi_\epsilon(y)$ (which is
always less than $1$), taking into account that $g\, \Phi_\epsilon =
K\, e^{-\Lambda} \Phi_\epsilon$, and taking the supremum over $(0,y)$,
\begin{equation}
  \label{eq:III-rig-4}
  N_\epsilon \leq   K(\delta) + N_\epsilon\, K_4\, \delta^{\lambda+1},
\end{equation}
As $\lambda+1 > 0$, taking $\delta$ small enough gives a bound for
$N_\epsilon$ which is independent of $\epsilon$, and hence proves that
$K$ is bounded on $(0,\delta)$. This in turn implies that it has a
strictly positive limit at $y=0$, as it is a nonincreasing function,
which can be seen from eqs. \eqref{eq:III-dmu}--\eqref{eq:III-def-h}.

\section{Partial uniqueness of scaling profiles}
\label{sec:uniqueness}

Let us prove theorems~\ref{thm:uniqueness-intro-II} and
\ref{thm:uniqueness-intro-III} on the partial uniqueness of scaling
profiles.  We will prove the following proposition, which already
contains theorem \ref{thm:uniqueness-intro-III}, and which will be
seen to easily imply theorem \ref{thm:uniqueness-intro-II} (see
section \ref{sec:global-uniqueness}):
\begin{prp}
  \label{prp:uniqueness}
  Assume that the coagulation coefficient $a$ is of the form
  \eqref{eq:kernel-expression-precise} with $\alpha \leq 0$. Assume
  that $g^1$ and $g^2$ are two self-similar profiles for Smoluchowski's
  equation, and also that
  \begin{enumerate}
  \item in the case $\alpha = 0$, $M_\lambda[g_1] = M_\lambda[g_2]$,
    and
    \begin{equation*}
      \lim_{y \to 0} y^{\tau-1} G^1(y)
      =
      \lim_{y \to 0} y^{\tau-1} G^2(y),
    \end{equation*}
    where $\tau$ is given by \eqref{eq:def-tau}, and $G^1$, $G^2$ are
    the primitives of $g^1$, $g^2$ based at $+\infty$, defined as in
    \eqref{eq:def-G}.
  \item in the case $\alpha < 0$, $M_\alpha[g_1] = M_\alpha[g_2]$,
    $M_\beta[g_1] = M_\beta[g_2]$, and
    \begin{equation*}
      \lim_{y \to 0} g^1(y) e^{\Lambda(y)}
      =
      \lim_{y \to 0} g^2(y) e^{\Lambda(y)},
    \end{equation*}
    where $\Lambda$ is given by \eqref{eq:III-Lambda}.
  \end{enumerate}
  Then, $g^1 = g^2$.
\end{prp}
We prove this proposition in two parts: in the next section we prove
that the result holds near $y=0$, and in section
\ref{sec:global-uniqueness} we prove that the result is global.

\subsection{Local result}

We will start by proving proposition \ref{prp:uniqueness} locally
near $y=0$; this is, we will show that under the same hypotheses there
exists $\delta > 0$ such that $g^1(y) = g^2(y)$ for $0 < y <
\delta$.

\subsubsection{Proof for $\alpha = 0$}

\paragraph{Step 1: Rewriting the equation for a difference of profiles.}

Assume that the coagulation coefficient $a$ is given by
\eqref{eq:kernel-expression-precise} with $\alpha = 0$ (so $\beta =
\lambda$). If $g$ is a self-similar profile, then
\eqref{eq:II-ss-rewrite}--\eqref{eq:def-h} from
lemma~\ref{lem:II-ss-rewrite} hold, so
\begin{equation}
  \label{eq:K(y)-2}
  K(y) = K_0 - \int_0^y z^{\tau-2} h(z) \,dz,
\end{equation}
with $K_0$ given by
\begin{equation}
  \label{eq:def-K_0}
  K_0 := \lim_{y \to 0} K(y) = \lim_{y \to 0} y^{\tau-1} G(y),
\end{equation}
which is known to exist and be strictly positive
\cite{citeulike:1271620}. Note that here we have integrated
eq. (\ref{eq:dK-in-lemma}) between $0$ and $y$, which can be done once
we know $K$ has a limit at $y=0$.

Gathering the above, we have
\begin{equation}
  \label{eq:ss-local-2}
  y^{\tau - 1} G(y) =
  K_0 -
  (1-\lambda)
  \int_0^y z^{\tau-2}
  \left(
    G * (y^\lambda g)
  \right)(z) \,dz,
\end{equation}
which is a remarkable equation in the sense that it is \emph{local
  near $0$}, and the term to the right is \emph{more regular near $0$}
than that to the left, as will be precised later. We also remark that
in \eqref{eq:ss-local-2} the parameters $\tau$ and $K_0$ depend on
$g$.

Now, let us obtain the corresponding equation for the difference of
two self-similar profiles. Let $g_1$, $g_2$ be two solutions of
self-similar profiles, and assume that they satisfy the conditions in
proposition~\ref{prp:uniqueness}. Then, both $g_1$ and $g_2$ satisfy
\eqref{eq:ss-local-2} with the same $\tau$ and $K_0$, and we can take
the difference to get
\begin{multline}
  \label{eq:ss-local-dif}
  y^{\tau - 1} \Delta G(y)
  =
  -
  \frac{1-\lambda}{2}
  \int_0^y z^{\tau-2}
  \left(
    (\Delta G) * (y^\lambda g_1)
  \right)(z) \,dz
  \\
  -
  \frac{1-\lambda}{2}
  \int_0^y z^{\tau-2}
  \left(
    G_2 * (y^\lambda \Delta g)
  \right)(z) \,dz,
\end{multline}
where
\begin{equation*}
  \Delta g := g_1 - g_2,
  \qquad
  \Delta G := G_1 - G_2.
\end{equation*}
Now, to obtain local uniqueness near $0$, take $\epsilon > 0$ and
write
\begin{equation}
  \label{eq:def-N}
  N
  \equiv
  N(\epsilon)
  := \sup_{y \in (0,\epsilon)} y^{\tau - 1} \abs{\Delta G(y)}.
\end{equation}
Observe that this quantity is known to be bounded thanks to
\cite{citeulike:1271620}. Let us prove from equation
\eqref{eq:ss-local-dif} that, if we take $\epsilon$ small enough, then
$N$ must be $0$.


\paragraph{Step 2: Estimate for the first term.}

Constants independent of $\epsilon$ will be denoted by $K_1$, $K_2$...
We will use the following bound, which holds for all solutions $g$ of
\eqref{eq:ss-profile-eq} \cite{citeulike:1271620} (and in particular
for $g_1$ and $g_2$):
\begin{equation}
  \label{eq:bound-g}
  g(y) \leq K_1 y^{-\tau}
  \quad \text{ for } y > 0
\end{equation}
for some constant $K_1 > 0$, which implies that
\begin{equation}
  \label{eq:bound-G}
  G(y) \leq K_2 y^{1-\tau}
  \quad \text{ for } y > 0,
\end{equation}
for some other constant $K_2$. Then, from \eqref{eq:bound-g},
\begin{multline}
  \label{eq:bound-1}
  \abs{ (\Delta G) *(y^\lambda g_1) (z) }
  \leq
  \int_0^z \abs{\Delta G(x)} (x-z)^\lambda \abs{g_1(x-z)} \,dx
  \\
  \leq
  K_1\, N
  \int_0^z x^{1-\tau} (x-z)^{\lambda-\tau} \,dx
  \leq
  K_3 \, N z^{2+\lambda-2\tau}.
\end{multline}
Hence, the first term on the right hand side of
\eqref{eq:ss-local-dif} can be estimated by
\begin{equation}
  \label{eq:bound-2}
  \abs{
    \int_0^y z^{\tau-2}
    \left(
      (\Delta G) * (y^\lambda g_1)
    \right)(z) \,dz
  }
  \leq
  N \, K_3
  \int_0^y z^{\lambda-\tau} \,dz
  =
  N\, K_4\, z^{1+\lambda-\tau}.
\end{equation}

\paragraph{Step 3: Estimate for the second term.}

For the second term in \eqref{eq:ss-local-dif} we need to make $\Delta
G$ appear instead of $\Delta g$. We use integration by parts to write:
\begin{multline}
  \label{eq:parts-1}
  \int_0^y z^{\tau-2} \left( G_2 * (y^\lambda \Delta g) \right)(z) \,dz
  \\
  =
  y^{\tau-2} \int_0^y \left( G_2 * (y^\lambda \Delta g) \right)(z) \,dz
  \\
  +
  (2-\tau) \int_0^y z^{\tau-3}
  \int_0^z \left( G_2 * (y^\lambda \Delta g) \right)(x) \,dx
  \,dz.
\end{multline}
Here, the boundary term at $y=0$ in the integration by parts vanishes,
which is a consequence of the bound below in eq. \eqref{eq:bound-3},
which we will show next for the term $\int_0^y G_2 * (y^\lambda \Delta
g)$ appearing above. Write:
\begin{multline}
  \label{eq:parts-2}
  \int_0^y \left( G_2 * (y^\lambda \Delta g) \right)(z) \,dz
  =
  D^{-1} \left( G_2 * (y^\lambda \Delta g) \right)(z) \,dz
  \\
  =
  (G_2 * D^{-1} (y^\lambda \Delta g)) (y).
\end{multline}
Also,
\begin{equation}
  \label{eq:parts-3}
  D^{-1} (y^\lambda \Delta g) (y)
  =
  - y^\lambda \Delta G(y) + \lambda \int_0^y z^{\lambda-1} \Delta G(z) \,dz,
\end{equation}
so
\begin{multline}
  \label{eq:parts-4}
  \abs{  D^{-1} (y^\lambda \Delta g) (y) }
  \leq
  \abs{ y^\lambda \Delta G(y)}
  + \lambda \int_0^y z^{\lambda-1} \abs{\Delta G(z)} \,dz
  \\
  \leq
  N y^{1-\tau+\lambda}
  + N \lambda  \int_0^y z^{\lambda-\tau} \,dz
  \leq
  K_5\, N\, y^{1+\lambda-\tau}.  
\end{multline}
Hence, from \eqref{eq:parts-2},
\begin{multline}
  \label{eq:parts-5}
  \abs{ \int_0^y \left( G_2 * (y^\lambda \Delta g) \right)(z) \,dz }
  \leq
  \abs{ (G_2 * D^{-1} (y^\lambda \Delta g)) (y) }
  \\
  \leq
  N\,K_2 \,K_5\, \int_0^y z^{1-\tau} (y-z)^{1+\lambda-\tau} \,dz
  \leq
  N\, K_6 \, y^{3+\lambda-2\tau}.
\end{multline}
And finally, gathering \eqref{eq:parts-1} and \eqref{eq:parts-5},
\begin{multline}
  \label{eq:bound-3}
  \abs{
    \int_0^y z^{\tau-2} \left( G_2 * (y^\lambda \Delta g) \right)(z) \,dz
  }
  \\
  \leq
  N\, K_6\, y^{\tau - 2} y^{3+\lambda - 2\tau}
  +
  N\, K_6\, \abs{2-\tau}
  \int_0^y z^{\tau-3} z^{3+\lambda - 2\tau} \,dz
  \\
  \leq
  N K_7\, y^{1+\lambda-\tau}.
\end{multline}

\paragraph{Step 4: Final estimate}

Now, from eq. \eqref{eq:ss-local-dif}, taking the supremum on
$(0,\epsilon)$ and using \eqref{eq:bound-2} and \eqref{eq:bound-3} one
has
\begin{equation}
  \label{eq:bound-4}
  N(\epsilon)
  \leq
  N(\epsilon) \,K_8\, \epsilon^{1+\lambda-\tau}.
\end{equation}
Note that $1 + \lambda - \tau > 0$ \cite{citeulike:1271620}, which is
crucial for this argument, and is a particular property of the
coagulation kernel we are using. Hence, for $\epsilon > 0$ small
enough, we have that
\begin{equation*}
  N \equiv N(\epsilon) = 0,
\end{equation*}
and hence that
\begin{equation*}
  g_1(y) = g_2(y)
  \quad
  \text{ for } 0 < y < \epsilon,
\end{equation*}
which proves the equality in proposition~\ref{prp:uniqueness} locally
near $y=0$ in the case $\alpha = 0$.

\subsubsection{Proof for $\alpha < 0$}

Assume again that $a$ is of the form
\eqref{eq:kernel-expression-precise}, now with $\alpha < 0$. Take two
self-similar solutions $g_1$, $g_2$ in the conditions of
proposition~\ref{prp:uniqueness}. Then lemma~\ref{lem:III-ss-rewrite}
applies to both $g_1$ and $g_2$ \emph{with the same $\Lambda$}, and
following the same reasoning as in the $\alpha = 0$ case we have
\begin{equation}
  \label{eq:III-uni-4}
  \Delta g (y)
  =
  \Delta K (y) e^{-\Lambda(y)},
\end{equation}
with
\begin{gather}
  \label{eq:III-uni-dK-2}
  (\Delta K)'(y)
  = \frac{1}{y} e^{\Lambda(y)} \Delta h(y)
  \quad
  \text{ for almost all } y > 0,
  \\
  \label{eq:III-def-h-2}
  \begin{split}
    \Delta h = &-(1-\lambda) (y^\alpha \Delta g) * (y^\beta g_2)
    \\
    &-(1-\lambda) (y^\alpha g_2) * (y^\beta \Delta g).
  \end{split}
\end{gather}
Here we denote $\Delta A := A_1 - A_2$ for any function $A$, and
$K_1,h_1,K_2,h_2$ are the functions $K,h$ associated to $g_1$, $g_2$ as
in lemma~\ref{lem:III-ss-rewrite}.

Then, from the equality of the limits in \eqref{eq:III-limits-equal}
we see that $\Delta K(y) \to 0$ as $y \to 0$, so integrating
\eqref{eq:III-uni-dK-2},
\begin{equation}
  \label{eq:III-uni-5}
  \Delta K (y)
  = \int_0^y \frac{1}{z} e^{\Lambda(z)} \Delta h(z) \,dz.
\end{equation}
Let us find a bound for $\Delta h$ using $\Delta K$. Take $\epsilon > 0$
and call
\begin{equation}
  \label{eq:III-uni-def-N}
  N := \sup_{y \in (0,\epsilon)} \abs{ \Delta K(y) }.
\end{equation}
Then, for the first term in
\eqref{eq:III-def-h-2},
\begin{multline}
  \label{eq:III-uni-6}
  \abs{ (y^\alpha \Delta g) * (y^\beta g_2) }
  \leq
  N \, (y^\alpha e^{-\Lambda(y)}) * (y^\beta g_2)
  \\
  \leq
  N K_0\, e^{-\Lambda(y)}\,
  (y^\beta)*(y^\alpha)
  \leq
  N K_1\,
  e^{-\Lambda(y)}\,
  y^{\lambda + 1},
\end{multline}
where we have also used that $g$ is bounded on $(0,\epsilon)$; of
course, we have much stronger information on its behavior at $y=0$,
which has been used through the fact that $\Delta K$ is bounded and
has a limit at $y=0$.

For the second term in \eqref{eq:III-def-h-2} a similar calculation
shows that
\begin{equation}
  \label{eq:III-uni-7}
  \abs{ (y^\alpha g_2) * (y^\beta \Delta g) }
  \leq
  N K_2\,
  y^{\lambda + 1}
  e^{-\Lambda(y)}.
\end{equation}
Putting together \eqref{eq:III-uni-6} and \eqref{eq:III-uni-7} we
obtain a bound for $\Delta h$:
\begin{equation}
  \label{eq:III-uni-bound-h}
  \abs{ \Delta h(y)}
  \leq
  N K_3\, y^{\lambda+1} e^{-\Lambda(y)},
\end{equation}
and continuing from \eqref{eq:III-uni-5},
\begin{equation}
  \label{eq:III-uni-9}
  \abs{ \Delta K (y) }
  \leq
  N K_3 \int_0^y z^{\lambda}  \,dz
  =
  N K_4\, y^{\lambda+1}.
\end{equation}
Finally, taking the supremum on $(0,\epsilon)$,
\begin{equation}
  \label{eq:III-uni-10}
  N
  \leq
  N K_4\, \epsilon^{\lambda+1},
\end{equation}
and then (as $\lambda+1 > 0$) choosing $\epsilon$ small enough proves
that $N = 0$. Hence,
\begin{equation}
  \label{eq:III-uni-final}
  g_1(y) = g_2(y)
  \quad \text{ for } y \in (0,\epsilon).
\end{equation}

\subsection{Global result}
\label{sec:global-uniqueness}

We now extend the local result in the previous section in order to
finish the proof of proposition \ref{prp:uniqueness}. After this it
will also be easy to prove theorem~\ref{thm:uniqueness-intro-II}.



\begin{proof}[End of proof of proposition \ref{prp:uniqueness}]
  Take two self-similar profiles $g_1$ and $g_2$ satisfying the
  hypotheses of the theorem. The results in previous sections show
  that $g_1 = g_2$ on some interval $(0,\epsilon)$, for some $\epsilon
  > 0$. Following a strategy usual in uniqueness theorems for ordinary
  differential equations, we will show that whenever $g_1 = g_2$ on an
  interval $(0,y_0)$, there is a $\delta > 0$ such that $g_1 = g_2$ on
  $(0, y_0+\delta)$. Together with our local uniqueness result, a
  well-known argument then shows that $g_1 = g_2$ on $(0,\infty)$.

  So, assume that $g_1 = g_2$ on $(0,y_0)$ for some $y_0 > 0$. Take
  $\delta > 0$ (to be fixed later) and define
  \begin{align}
    \label{eq:global-extension-N}
    N &:= \sup_{y \in (y_0, y_0+\delta)} \abs{\Delta g(y)}
    \\
    & = \sup_{y \in (0, y_0+\delta)} \abs{\Delta g(y)}.
  \end{align}
  Choosing $\delta$ appropriately, we will prove that $N = 0$. Take
  the difference of the self-similar equation for $g_1$ and $g_2$ to
  get
  \begin{equation}
    \label{eq:global-extension-0}
    y \Delta g
    =
    \Delta G + (1-\lambda) D_{-1} C(\Delta g, g_1 + g_2).
  \end{equation}
  Hence, taking the supremum on $(y_0, y_0+\delta)$,
  \begin{multline}
    \label{eq:global-extension-0.1}
    y_0\, N
    \leq
    \norm{\Delta G}_{L^\infty(y_0,y_0+\delta)}
    \\
    + (1-\lambda)
    \norm{ D_{-1} C(\Delta g, g_1 + g_2) }_{L^\infty(y_0,y_0+\delta)}.
  \end{multline}
  To bound the first term we have, for $y \in (y_0, y_0+\delta)$,
  \begin{equation}
    \label{eq:global-extension-0.2}
    \abs{\Delta G(y)}
    = \abs{ \int_{y_0}^{y} \delta g(z) \,dz }
    \leq
    N (y-y_0)
    \leq
    N (\delta-y_0).
  \end{equation}
  The bound for the second term in \eqref{eq:global-extension-0.1}
  depends on the type of kernel we are considering:

  \paragraph{Bound for an $\alpha < 0$ kernel}
  
  \begin{multline}
    \label{eq:global-extension-1}
    2 C(\Delta g, g_1)
    = \{y^\alpha \Delta g \}*\{ y^\beta g_1 \}
    + \{y^\beta \Delta g\}*\{y^\alpha g_1 \}
    \\
    =
    (y^\alpha \Delta g ) * \{ y^\beta g_1 \}
    + (y^\beta \Delta g) * \{y^\alpha g_1 \},
  \end{multline}
  so, for $\delta \leq y \leq 2 \delta$,
  \begin{multline*}
    \label{eq:global-extension-1.5}
    \abs{C(\Delta g, g_1)}
    \\
    \leq
    \norm{y^\alpha \Delta g}_{L^\infty(0,y)}
    M_\beta
    +
    \norm{y^\beta \Delta g}_{L^\infty(0,y)}
    M_\alpha
    \leq
    K_0 N,
  \end{multline*}
  where the first inequality has been obtained by applying Young's
  inequality for measures (or, alternatively, by writing out
  explicitly $C(\Delta g, g_1)$ with its classical formula and
  bounding each term).
  Hence,
  \begin{equation}
    \label{eq:global-extension-3}
    \int_\delta^y \abs{C(\Delta g, g_1)} \,dz
    \leq
    K_0 N \,(y - \delta).
  \end{equation}

  \paragraph{Bound for an $\alpha = 0$ kernel}
  
  \begin{multline}
    \label{eq:global-extension-4}
    2 C(\Delta g, g_1)
    = \{y^\lambda \Delta g \}*\{ g_1 \}
    + \{\Delta g\}*\{y^\lambda g_1 \}
    \\
    =
    (y^\lambda \Delta g ) * \{ g_1 \}
    + \{ \Delta g \} * (y^\lambda g_1 )
    - M_\lambda \{ \Delta g \},
  \end{multline}
  and then
  \begin{multline}
    \label{eq:global-extension-5}
    2 \abs{D_{-1} C(\Delta g, g_1)}
    \\
    \leq
    D_{\lambda-1}(y^\lambda \Delta g ) * D_{-\lambda} \{ g_1 \}
    + D_{-1} \{ \Delta g \} * (y^\lambda g_1 )
    + M_\lambda D_{-1} \{ \Delta g \}
    \\
    \leq
    M_\lambda \norm{ D_{\lambda-1}(y^\lambda \Delta g )}_{L^\infty(-\infty,y)}
    + 2 M_\lambda (y-\delta) N.
  \end{multline}
  We need to bound $\norm{ D_{\lambda-1}(y^\lambda \Delta g
    )}_{L^\infty(-\infty,y)}$:
  \begin{multline}
    \label{eq:global-extension-6}
    D_{\lambda-1}(y^\lambda \Delta g)
    =
    \frac{1}{\Gamma(1-\lambda)}
    \int_{-\infty}^y z^\lambda \Delta g(z) (y-z)^{-\lambda} \,dz
    \\
    \frac{1}{\Gamma(1-\lambda)}
    \int_{\delta}^y z^\lambda \Delta g(z) (y-z)^{-\lambda} \,dz
    \\
    \leq
    K_1 N
    \int_{\delta}^y (y-z)^{-\lambda} \,dz
    \\
    =
    K_1 N \frac{1}{1-\lambda}
    (y-\delta)^{1-\lambda}
    .
  \end{multline}

  Finally, with \eqref{eq:global-extension-3} and
  \eqref{eq:global-extension-6} we can continue from
  \eqref{eq:global-extension-0.1} to get
  \begin{equation}
    \label{eq:global-extension-7}
    y_0\, N
    \leq
    K_2\, N\, (y-\delta)^{1-\lambda},
  \end{equation}
  so choosing $\delta$ small enough we can deduce that $N=0$. This
  finishes the proof.
\end{proof}

Let us finally prove theorem~\ref{thm:uniqueness-intro-II}:

\begin{proof}[Proof of theorem~\ref{thm:uniqueness-intro-II}]
  Assume that $a$ is of the form \eqref{eq:kernel-expression-precise}
  with $\alpha = 0$, and take two solutions $g_1$, $g_2$ of equation
  \eqref{eq:ss-profile-eq} in the conditions of
  theorem~\ref{thm:uniqueness-intro-II}, this is,
  \begin{gather*}
    \int_0^\infty y\, g_1(y) \,dy
    =
    \int_0^\infty y\, g_2(y) \,dy,
    \\
    \int_0^\infty y^\lambda\, g_1(y) \,dy
    =
    \int_0^\infty y^\lambda\, g_2(y) \,dy.
  \end{gather*}
  Then, for any $\mu > 0$, the function
  \begin{equation*}
    \tilde{g}_1(y) := \mu^{1+\lambda} g_1(\mu y)
    \quad \text{ for } y > 0
  \end{equation*}
  is another solution of equation \eqref{eq:ss-profile-eq}, which is a
  simple consequence of the homogeneity of the coagulation coefficient
  $a$ (see, for example, \cite{citeulike:1271620}). We also check
  easily that
  \begin{gather}
    \int_0^\infty y^\lambda\, \tilde{g}_1(y) \,dy
    =
    \int_0^\infty y^\lambda\, g_1(y) \,dy
    \\
    \label{eq:mass-scaling}
    \int_0^\infty y\, \tilde{g}_1(y) \,dy
    =
    \mu^{\lambda-1} \int_0^\infty y\, g_1(y) \,dy
    \\
    \lim_{y \to 0} y^{\tau - 1} \int_y^\infty \tilde{g}_1(z) \,dz
    =
    \mu^{1+\lambda-\tau} \lim_{y \to 0} y^{\tau - 1} \int_y^\infty g_1(z) \,dz,
  \end{gather}
  where $\tau := 2 - (1-\lambda) M_\lambda[g_1]$. Hence, the moment of
  order $\lambda$ of $\tilde{g}_1$ is the same no matter which $\mu$
  we take, so we can choose $\mu > 0$ in such a way that
  \begin{gather*}
    \int_0^\infty y^\lambda\, \tilde{g}_1(y) \,dy
    =
    \int_0^\infty y^\lambda\, g_2(y) \,dy
    \\
    \lim_{y \to 0} y^{\tau - 1} \int_y^\infty \tilde{g}_1(z) \,dz
    =
    \lim_{y \to 0} y^{\tau - 1} \int_y^\infty g_2(z) \,dz.
  \end{gather*}
  Hence, with this value of $\mu$, by proposition~\ref{prp:uniqueness}
  we have that $\tilde{g}_1 = g_2$, and in particular their masses are
  equal; as the masses of $g_1$ and $g_2$ are also equal, from
  \eqref{eq:mass-scaling} we deduce that in fact $\mu$ must be equal
  to $1$, so $g_1 = \tilde{g}_1 = g_2$, which shows the result.
\end{proof}

\section{Appendix: fractional derivatives}
\label{sec:appendix}

The extension of the concept of integration and differentiation to
include derivatives and integrals of noninteger order is a well
established theory
\cite{citeulike:1915345,citeulike:1428686,citeulike:1428685}. Here we
give a brief but self-contained introduction to it and state without
proof the main standard results. Some particular properties needed in
the rest of this paper and which are not commonly encountered will be
given with complete proofs below.

For our purposes, the simplest and most general definition of
fractional derivatives is given in the context of distributions, and
can be found in the book by Schwartz
\cite[VI.5]{citeulike:1915345}. The reader can check that our
definitions are the same as those given there, even if the
presentation is somewhat different. Other expositions are found in
\cite{citeulike:1428686,citeulike:1428685}, and we refer to those
sources for the proof of the main results on fractional
differentiation given below.

In the following we will use the space $\mathcal{C}^\infty_R$
consisting of all infinitely differentiable functions $f:\RR \to \RR$
which have support bounded below; this is, those $f$ which have
support contained in $[a,+\infty)$ for some $a \in \RR$. The notation
$\mathcal{C}^\infty_R$ is intended to suggest that the important part
of a function $f$ is \emph{to the right}, if one represents the real
line as usual.\footnote{We prefer to write $\mathcal{C}^\infty_R$
  instead of $\mathcal{D}_+$, used in Schwartz's book, as in the
  literature related to Smoluchowski's equation the subscript `$+$' is
  frequently used to denote that functions in the involved space are
  nonnegative, which could cause confusion here, where we want to
  stress a property of \emph{their support}.} Analogously, we define
$\mathcal{C}^\infty_L$ as the set of all infinitely differentiable
functions $f: \RR \to \RR$ whose support is bounded above.

We will first define fractional derivatives for smooth functions, and
then extend the concept to distributions with a common duality method.

\subsection{Fractional derivatives of smooth functions}
\label{sec:fd-smooth}

\begin{dfn}[Left fractional derivatives]
  \label{dfn:left-derivative}
  For $f \in \mathcal{C}^\infty_R$ and real $k > 0$, we define
  \emph{the left integral or order $k$ of $f$} as
  \begin{equation}
    \label{eq:fractional-integral}
    D^{-k} f(y)
    :=
    \frac{1}{\Gamma(k)} \int_{-\infty}^y f(z) (y-z)^{k-1} \,dz
    \quad
    \text{ for } y \in \RR,
  \end{equation}
  where $\Gamma$ is the Gamma function.
  For $k = 0$ we just write $D^k f = f$. For real $k \geq 0$ we write
  $k$ as $k = n-s$, with $n > 0$ an integer and $0 \leq s < 1$, and
  define \emph{the left derivative of order $k$ of $f$} to be
  \begin{equation}
    \label{eq:fractional-derivative}
    D^k f := \frac{d^n}{dy^n} (D^{-s} f).
  \end{equation}
\end{dfn}
The above is a usual definition of fractional integrals and
derivatives \cite{citeulike:1428686}, sometimes called \emph{the
  Riemann-Lebesgue definition}. Names given to the above also differ
slightly from place to place: we may refer to $D^k f$ as \emph{the
  left derivative of order $k$ of $f$}, or just \emph{the $k$-th
  derivative of $f$}, for any real $k$ (even for $k < 0$), thus
emphasizing that all $D^k$ are part of a family of operators with
common properties.

There is a completely analogous concept of \emph{right derivative}
where integrals are taken \emph{from $+\infty$}:

\begin{dfn}[Right fractional derivatives]
  \label{dfn:right-derivative}
  For $f \in \mathcal{C}^\infty_L$ and real $k > 0$, we define
  \emph{the right integral or order $k$ of $f$} as
  \begin{equation}
    \label{eq:fractional-integral-right}
    D_{-k} f(y)
    :=
    \frac{1}{\Gamma(k)} \int_y^\infty f(z) (z-y)^{k-1} \,dz
    \quad
    \text{ for } y \in \RR,
  \end{equation}
  where $\Gamma$ is the Gamma function.
  For $k = 0$ we just write $D_k f = f$. For real $k \geq 0$ we write
  $k$ as $k = n-s$, with $n > 0$ an integer and $0 \leq s < 1$, and
  define \emph{the left derivative of order $k$ of $f$} to be
  \begin{equation}
    \label{eq:fractional-derivative-r}
    D_k f := (-1)^n \frac{d^n}{dy^n} (D_{-s} f).
  \end{equation}
\end{dfn}

Some easy consequences are the following:
\begin{enumerate}
\item For $k \in \RR$ and $f \in \mathcal{C}^\infty_R$, $D^k f$ is
  again on $\mathcal{C}^\infty_R$, and the analogous result holds for
  $D_k$ and $\mathcal{C}^\infty_L$.
\item For integer $k \geq 0$, $D^k$ is just the usual $k$-th
  derivative of $f$, while $D^{-k}$ is the $k$-fold iteration of the
  primitive based at $-\infty$; in particular, $D^{-1}f$ is the only
  primitive of $f$ which is $0$ at $-\infty$. Analogously, $D_k$ is
  the usual $k$-th derivative of $f$, \emph{times $(-1)^k$} (see next
  for a natural reason for this definition), while $D_{-k}$ is the
  $k$-fold iteration of the primitive based at $+\infty$.
\item Right derivatives are the concept symmetric to that of left
  derivatives under the reflection of $\RR$: if we define \emph{the
    reflection} of a function $f: \RR \to \RR$ to be the function $\mathcal{R}f:
  \RR \to \RR$ given by $\mathcal{R}f(y) := f(-y)$, then
  \begin{equation}
    \label{eq:right_derivative-reflection}
    D_k(f) := \mathcal{R} (D^k (\mathcal{R}f))
    \quad
    \text{ for } k \in \RR, f \in \mathcal{C}^\infty_L.
  \end{equation}
  Alternatively, one can take this as a definition of right
  derivatives from the perhaps more natural concept of left
  derivatives. Note that the alternating sign in equation
  \eqref{eq:fractional-derivative-r} is unavoidable if we want to
  conserve this symmetry property.
\end{enumerate}
Remarkably, the following composition result holds:

\begin{thm}
  \label{thm:D_can_be_composed}
  For real $k,j$,
  \begin{gather}
    \label{eq:D_can_be_composed}
    D^j (D^k f) = D^{j+k} f
    \quad
    \text{ for any } f \in \mathcal{C}^\infty_R,
    \\
    \label{eq:D_can_be_composed-right}
    D_j (D_k f) = D_{j+k} f
    \quad
    \text{ for any } f \in \mathcal{C}^\infty_L.
  \end{gather}
\end{thm}
A proof follows from elementary analysis arguments. For this result to
hold it is essential that our definitions \ref{dfn:left-derivative}
and \ref{dfn:right-derivative} above have picked specific primitives
(those which are $0$ at $-\infty$ or $+\infty$, respectively) out of
all the possible primitives of a function $f$. Said in another way,
the spaces $C^\infty_R$, $C^\infty_L$ in which we are working only
contain \emph{one} of all the possible primitives of a given function
$f$, and thus the above composition rule can hold.

Also, it is easy to see that $D_k$ is the dual of $D^k$ in the
following sense:
\begin{lem}
  For $f \in \mathcal{C}^\infty_R$ and $g \in
  \mathcal{C}^\infty_L$ it holds that
  \begin{equation}
    \label{eq:Dk-D_k-dual}
    \int_{-\infty}^{+\infty} D^k f(y)\, g(y) \,dy
    =
    \int_{-\infty}^{+\infty} f(y)\, D_k g(y) \,dy.
  \end{equation}
\end{lem}
This suggests the definition for distributions given below.



\subsection{Fractional derivatives of distributions}

Consider the set $\mathcal{D}_L'$ of distributions on $\RR$ which have
compact support \emph{bounded below}. One can show that
$\mathcal{D}_L'$ is the dual of $\mathcal{C}^\infty_L$ when the latter
is equipped with a natural topology (which extends that of
$\mathcal{C}^\infty_0$) \cite[VI.5]{citeulike:1915345}. So,
$\mathcal{D}_L'$ should be thought of as $(\mathcal{C}^\infty_L)'$,
which is useful for remembering that distributions in this space have
support contained in $(a,\infty)$ for some $a \in \RR$. For these
distributions, one can define $\ap{T, \psi}$ for any $\psi \in
\mathcal{C}^\infty_L$ as
\begin{equation}
  \label{eq:duality-product-noncompact-support}
  \ap{T, \psi}
  :=
  \langle T, \tilde{\psi} \rangle
\end{equation}
for any $\tilde{\psi} \in \mathcal{C}^\infty_0(\RR)$ which coincides
with $\psi$ on the support of $T$. Of course, this definition does not
depend on the particular extension chosen. We define $\mathcal{D}'_L$
analogously, and also the pairing $\ap{T, \phi}$ for any $T \in
\mathcal{D}'_L$, $\phi \in \mathcal{C}^\infty_R$.

\begin{dfn}[Fractional derivatives of distributions]
  Take $k \in \RR$. For a distribution $T \in \mathcal{D}'_L$ we
  define the distribution $D^k T$ as
  \begin{equation}
    \label{eq:def_derivative_distribution-left}
    \langle  D^k T, \psi  \rangle
    := \langle  T, D_k \psi  \rangle
    \quad \text{ for } \psi \in \mathcal{C}^\infty_L(\RR).
  \end{equation}
  Analogously, for a distribution $T \in \mathcal{D}'_R$ we
  define the distribution $D_k T$ as
  \begin{equation}
    \label{eq:def_derivative_distribution-right}
    \langle  D_k T, \psi  \rangle
    := \langle  T, D^k \phi  \rangle
    \quad \text{ for } \phi \in \mathcal{C}^\infty_L(\RR).
  \end{equation}
  Here, $D^k \phi$ and $D_k \psi$ are the right and left fractional
  derivatives, respectively, defined in section
  \ref{sec:fd-smooth}. Note that the duality products here are well
  defined as indicated in
  (\ref{eq:duality-product-noncompact-support}). Also, this agrees
  with definitions \ref{dfn:left-derivative} and
  \ref{dfn:right-derivative} when $T$ is a function in
  $\mathcal{C}^\infty_R$ (or $\mathcal{C}^\infty_L$), as can be seen
  from (\ref{eq:Dk-D_k-dual}).
\end{dfn}
Then, the $D^k$ are linear operators for which the composition rule
(\ref{eq:D_can_be_composed}) still holds: for any $j,k \in \RR$,
\begin{gather}
  \label{eq:D_can_be_composed-distributions}
  D^j (D^k T) = D^{j+k} T
  \quad
  \text{ for any } T \in \mathcal{D}'_R
  \\
  D_j (D_k T) = D_{j+k} T
  \quad
  \text{ for any } T \in \mathcal{D}'_L.
\end{gather}

The convolution of two distributions in $\mathcal{D}_R'$ (or
two distributions in $\mathcal{D}_L'$) is well defined
\cite[VI.5]{citeulike:1915345}, as we show below:

\begin{dfn}[Convolution of a distribution and a smooth function]
  Given $T \in \mathcal{D}'_R$ and $\phi \in \mathcal{C}^\infty_L$ we
  define $T * \phi$ as the distribution in $\mathcal{D}'_R$ given by
  \begin{equation}
    \label{eq:def-convolution-distribution-phi}
    \ap{ T * \phi, \psi } := \ap{ T, (\mathcal{R}\phi)*\psi }
    \quad
    \text{ for } \psi \in \mathcal{C}^\infty_R.
  \end{equation}
  Note that $(\mathcal{R}\phi)*\psi \in \mathcal{C}^\infty_R$. For $T \in
  \mathcal{D}'_L$ and $\psi \in \mathcal{C}^\infty_R$ the convolution
  $T*\psi$ is defined analogously.
\end{dfn}

The convolution with a function in $\mathcal{C}^\infty_R$ (or
$\mathcal{C}^\infty_L$) is regularizing, as it happens in the more
familiar case of convolution with a $C^\infty$ function with bounded
support: if $T \in \mathcal{D}'_L$ and $\psi \in
\mathcal{C}^\infty_R$, then one can prove that $T * \psi$ is equal to
a function in $\mathcal{C}^\infty_R$, given by
\begin{equation}
  \label{eq:T*phi_is_regular}
  T*\psi (y) = \ap{T, \tau_y \psi}
  \quad
  \text{ for } y \in \RR,
\end{equation}
where $(\tau_y\psi)(x) := \psi(x-y)$ is the translation of $\psi$ by
$y$. Similarly the convolution $T*\phi$ for $T \in \mathcal{D}_R'$ and
$\phi \in \mathcal{C}^\infty_L$ is a function in
$\mathcal{C}^\infty_L$.

\begin{dfn}[Convolution of two distributions]
  Given $T, S \in \mathcal{D}_R'$, we define $T*S$ as the distribution
  in $\mathcal{D}_R'$ given by
  \begin{equation}
    \label{eq:def-convolution-distributions}
    \ap{ T*S, \psi } := \ap{ T, (\mathcal{R}S) * \psi }
    \quad \text{ for } \psi \in \mathcal{C}^\infty_R.
  \end{equation}
  (As remarked before (\ref{eq:T*phi_is_regular}), $(\mathcal{R}S) *
  \psi \in \mathcal{C}^\infty_R$.) The convolution $T*S$ for $T, S \in
  \mathcal{D}_L'$ is defined analogously.
\end{dfn}

The following well-known result on the derivation of a convolution
holds in complete generality with these definitions:

\begin{thm}
  \label{thm:convolution-derivative-distributions}
  For any $T, S \in \mathcal{D}_L'$ and any $k \in \RR$,
  \begin{equation*}
    D^k(T*S) = (D^k T) * S.
  \end{equation*}
\end{thm}

Although fractional derivation and integration operators of
non-integer order are not local, their regularity properties
nevertheless depend only on local properties of the function they act
on, as can be easily proved by observing that $D^k f$ can be written
as a convolution of $f$ with a distribution which is
$\mathcal{C}^\infty$ away from $0$ \cite[VI.5, p.
174]{citeulike:1915345}. A manifestation of this is the following
result:

\begin{thm}
  Let $T$ be a distribution on $\RR$ with compact support to the left
  (this is, $T \in \mathcal{D}_L'$), and such that $T$ is
  $\mathcal{C}^\infty(U)$, for $U \subseteq \RR$ a given open set.
  Then, $D^k T \in \mathcal{C}^\infty(U)$ for all $k \in \RR$.
\end{thm}

It is well known that multiplying a distribution by a
$\mathcal{C}^\infty$ function preserves its local regularity, and this
is still true when one measures this regularity in terms of the
integrability of a given fractional derivative. As this result is not
easily found in the literature and its proof is not obvious, we give
it in section \ref{sec:product-regularity} below:

\begin{thm}
  \label{thm:local-regularity-of-product}
  Let $T$ be a distribution on $\RR$ with compact support to the left
  (this is, $T \in \mathcal{D}_L'$), and assume that $D^\mu T$ is
  locally integrable for some $\mu > 0$.  Then, for any smooth
  function $\Phi$ on $\RR$, $D^\mu (\Phi T)$ is locally integrable.
\end{thm}
Of course, the analogous result holds for a distribution with compact
support to the right and right derivatives $D_\mu$.

Let us also prove a result which is used in this paper, which says
that the property that $D^k T$ is locally integrable is stronger the
higher $k$ is:

\begin{lem}
  \label{lem:less-derivatives-better}
  Let $T \in \mathcal{D}_L'$ be a distribution on $\RR$ with compact
  support to the left, and assume that $D^k T$ is locally integrable
  on $\RR$ for some $k \in \RR$. Then, $D^{k-m} T$ is locally
  integrable for all real $m \geq 0$.
\end{lem}

\begin{proof}
  It is enough to prove it for $k = 0$, as then the general result is
  obtained by applying this particular case to the distribution $D^k
  T$, taking into account the composition law
  (\ref{eq:D_can_be_composed-distributions}).

  Then, to prove it for $k=0$, take $T$ a locally integrable function
  on $\RR$ with support contained on $(R, \infty)$. Fix a compact
  interval $[a,b]$ with $b > R$. For a test function $\phi \in
  \mathcal{C}^\infty_0(\RR)$ with compact support contained on $(a,b)$
  and any $m > 0$ we have
  \begin{equation*}
    \ap{D^{-m} T, \phi}
    =
    \frac{1}{\Gamma(m)}
    \int_{R}^b T(y)
    \int_y^b \phi(z) (z-y)^{m-1} \,dz
    \,dy,
  \end{equation*}
  and hence
  \begin{multline*}
    \abs{\ap{D^{-m} T, \phi}}
    \leq
    \norm{\phi}_\infty
    \frac{1}{\Gamma(m)}
    \int_{R}^b \abs{T(y)}
    \int_y^b (z-y)^{m-1} \,dz
    \,dy
    \\
    =
    \norm{\phi}_\infty
    \frac{1}{\Gamma(m)}
    \int_{R}^b \abs{T(y)}
    \int_0^{b-y} z^{m-1} \,dz
    \,dy
    \\
    \leq
    \norm{\phi}_\infty
    \frac{(b-R)^{m}}{m \Gamma(m)}
    \int_{R}^b \abs{T(y)}
    \,dy,
  \end{multline*}
  which proves that $D^{-m}T$ is locally integrable on $(a,b)$.
\end{proof}

\subsection{Regularity of a product by a smooth function}
\label{sec:product-regularity}

In this section we give the proof of
theorem~\ref{thm:local-regularity-of-product} on the regularity of
fractional order of a product by a $\mathcal{C}^\infty$ function. We
recall its statement:

\begin{thm}
  Let $T \in \mathcal{D}_L'$ be a distribution on $\RR$ with compact
  support to the left, and assume that $D^\mu T$ is locally integrable
  for some $\mu > 0$.  Then, for any smooth function $\Phi$ on $\RR$,
  $D^\mu (\Phi T)$ is locally integrable.
\end{thm}

The proof is broken into several lemmas. The following one sometimes
serves as a weaker substitute for the rule of differentiation of a
product:
\begin{lem}
  \label{lem:weak-product-rule}
  Take $0 < k < 1$. If $\phi, \psi \in \mathcal{C}^\infty(\RR)$ and
  have compact support to the right, then the following equality holds:
  \begin{multline*}
    D_k (\phi D_{-k} \psi) (y)
    =
    \phi(y) \psi(y)
    \\
    - \frac{\sin (\pi k)}{\pi}
    \int_y^\infty \psi(x)
    \frac{1}{x-y} \int_y^x \phi'(z) (x-z)^{k} (z-y)^{-k} \,dz
    \,dx.
  \end{multline*}
\end{lem}

\begin{proof}
  We have $D_k (\phi D_{-k} \psi) = D_1 D_{k-1} (\phi D_{-k}
  \psi)$. Let us calculate this:
  \begin{multline}
    \label{eq:ld-1}
    D_{k-1} (\phi D_{-k} \psi) (y)
    \\
    =
    \frac{1}{\Gamma(k) \Gamma(1-k)}
    \int_y^\infty \phi(z)
    \int_z^\infty \psi(x) (x-z)^{k-1}
    (z-y)^{-k} \,dx \,dz
    \\
    =
    K_1
    \int_y^\infty \psi(x)
    \int_y^x \phi(z) (x-z)^{k-1}
    (z-y)^{-k} \,dz \,dx,
  \end{multline}
  where we have set $K_1 := 1/(\Gamma(k) \Gamma(1-k))$ for short.  By
  a change of variables $u = (z-y)/(x-y)$, the inner integral can be
  written as
  \begin{equation*}
    \int_y^x \phi(z) (x-z)^{k-1} (z-y)^{-k} \,dz
    = \int_0^1 \phi(u(x-y) + y) (1-u)^{k-1} u^{-k} \,du,
  \end{equation*}
  so we can calculate the derivative in $y$ of \eqref{eq:ld-1} and
  obtain
  \begin{multline*}
    D_k (\phi D_{-k} \psi) (y)
    =
    - K_1
    \frac{d}{dy}
    \int_y^\infty \psi(x)
    \int_y^x \phi(z) (x-z)^{k-1}
    (z-y)^{-k} \,dz \,dx
    \\
    =
    - K_1
    \frac{d}{dy}
    \int_y^\infty \psi(x)
    \int_0^1 \phi(u(x-y) + y) (1-u)^{k-1} u^{-k} \,du
    \,dx
    \\
    =
    \frac{B(k, 1-k)}{\Gamma(k) \Gamma(1-k)}
    \psi(y) \phi(y)
    \\
    -K_1
    \int_y^\infty \psi(x)
    \int_0^1 \phi'(u(x-y) + y) (1-u)^k u^{-k} \,du
    \,dx
    \\
    =
    \psi(y) \phi(y)
    -K_1
    \int_y^\infty \psi(x)
    \frac{1}{x-y}
    \int_0^1 \phi'(z) (x-z)^k (z-y)^{-k} \,dz
    \,dx,
  \end{multline*}
  where $B(k, 1-k)$ is the Beta function for the parameters
  $(k,1-k)$. We have used the well-known relationship between the Beta
  and Gamma functions, and have undone our previous change of
  variables. This is the expression in the lemma; note that $K_1$ is
  the constant that appears there.
\end{proof}

\begin{lem}
  Take $k \geq 0$ and $\Phi \in \mathcal{C}^\infty_b(\RR)$. Then, for
  all $\psi \in \mathcal{C}^\infty_0(\RR)$ with compact support
  contained on a fixed interval $(-\infty, b)$ it holds that
  \begin{equation}
    \label{eq:simpler-ineq}
    \abs{ D_{-k} (\Phi D_{k-1} \psi) (y) }
    \leq
    (b-y)
    \norm{\psi}_\infty
    \norm{\Phi}_\infty
    \quad \text{ for all } y < b.
  \end{equation}
\end{lem}

\begin{proof}
  For $y \geq b$, $D_{-k} (\Phi D_{k-1} \psi) (y)$ is zero. For $y < b$ we have
  \begin{multline*}
    \abs{ D_{-k} (\Phi D_{k-1} \psi) (y) }
    \\
    \leq
    \frac{1}{\Gamma(k) \Gamma(1-k)}
    \int_y^b \abs{\Phi(z)} (z-y)^{k-1}
    \int_z^b \abs{\psi(x)} (x-z)^{-k}
    \,dx \,dz
    \\
    \leq
    \frac{1}{\Gamma(k) \Gamma(1-k)}
    \norm{\psi}_\infty
    \norm{\Phi}_\infty
    \int_y^b 
    \int_y^x (z-y)^{k-1} (x-z)^{-k}
    \,dz \,dx
    \\
    \leq
    \frac{B(k,1-k)}{\Gamma(k) \Gamma(1-k)}
    (b-y)
    \norm{\psi}_\infty
    \norm{\Phi}_\infty
    =
    (b-y)
    \norm{\psi}_\infty
    \norm{\Phi}_\infty,
  \end{multline*}
  taking into account the relationship between the Gamma and Beta
  functions.
\end{proof}

\begin{lem}
  \label{lem:bounded_operator-previous}
  Take $\Phi \in \mathcal{C}^\infty_b(\RR)$ and $0 \leq k < 1$. Fix $a <
  b \in \RR$. Then, for all $\psi \in \mathcal{C}^\infty_0(\RR)$ with
  compact support contained on a fixed interval $(-\infty, b)$ it
  holds that
  \begin{equation*}
    \norm{ D_{k} ( \Phi D_{-k} \psi) }_{L^{\infty}(a,b)}
    \leq
    K \norm{\psi}_\infty,
  \end{equation*}
  where $K \geq 0$ is a constant that only depends on $k$, $\Phi$ and
  the interval $(a,b)$.
\end{lem}

\begin{proof}
  For $k=0$ the statement is trivial. For $0 < k < 1$, we use
  lemma~\ref{lem:weak-product-rule} to write, for $y < b$,
  \begin{multline*}
    D_{k} ( \Phi D_{-k} \psi) (y)
    =
    \Phi(y) \psi(y)
    \\
    - \frac{\sin (\pi k)}{\pi}
    \int_y^b \psi(x)
    \frac{1}{x-y} \int_y^x \Phi'(z) (x-z)^{k} (z-y)^{-k} \,dz
    \,dx,
  \end{multline*}
  where we have also taken into account that $\psi(x)$ is $0$ for $x
  \geq b$. Here, the first term has the straightforward bound
  $\norm{\Phi \psi}_\infty \leq \norm{\Phi}_\infty
  \norm{\psi}_\infty$. As for the second one,
  \begin{multline*}
    \abs{
      \int_y^b \psi(x)
      \frac{1}{x-y} \int_y^x \Phi'(z) (x-z)^{k} (z-y)^{-k} \,dz
      \,dx
    }
    \\
    \leq
    \norm{\psi}_\infty \norm{\Phi'}_\infty
    \int_y^b
    \frac{1}{x-y} \int_y^x (x-z)^{k} (z-y)^{-k} \,dz
    \,dx
    \\
    =
    K\, (b-y)  \norm{\psi}_\infty \norm{\Phi'}_\infty,
  \end{multline*}
  where the constant $K$ is $B(1+k,1-k)$. Using this for $a < y <
  b$ proves the statement.
\end{proof}

\begin{lem}
  \label{lem:bounded_operator}
  Take $\Phi \in \mathcal{C}^\infty_b(\RR)$ and $k < 1$. Fix $a <
  b \in \RR$. Then, for all $\psi \in \mathcal{C}^\infty_0(\RR)$ with
  compact support contained on a fixed interval $(-\infty, b)$ it
  holds that
  \begin{equation*}
    \norm{ D_{k} ( \Phi D_{-k} \psi) }_{L^{\infty}(a,b)}
    \leq
    K \norm{\psi}_\infty,
  \end{equation*}
  where $K \geq 0$ is a constant that only depends on $k$, $\Phi$ and
  the interval $(a,b)$.
\end{lem}

\begin{proof}
  Lemma~\ref{lem:bounded_operator-previous} proves this when $0 \leq k
  < 1$, and we can prove the general case inductively: if the result
  is valid for a given $k < 1$, then
  \begin{multline*}
    D_{k-1} ( \Phi D_{1-k} \psi)
    =
    D_{k-1} ( \Phi D_1 (D_{-k} \psi))
    \\
    =
    D_{k-1} D_1 ( \Phi D_{-k} \psi)
    -
    D_{k-1} ((D_1 \Phi) (D_{-k} \psi))
    \\
    =
    D_{k} ( \Phi D_{-k} \psi)
    -
    D_{k-1} ((D_1 \Phi) (D_{-k} \psi)).
  \end{multline*}
  For the first term we can use our induction hypothesis, and the
  second one can be bounded thanks to lemma~\eqref{eq:simpler-ineq}.
  This shows the lemma.
\end{proof}

\begin{proof}[Proof of theorem~\ref{thm:local-regularity-of-product}]
  Take $R \in \RR$ so that $T$ has support contained in $(R,
  \infty)$. Fix a compact interval $(a,b)$ with $b > R$. For a
  function $\psi \in \mathcal{C}^\infty_0(\RR)$ with compact support
  on $(a,b)$ we have
  \begin{multline*}
    \ap{D^\mu (\Phi T), \psi}
    =
    \ap{\Phi T, D_{\mu} \psi}
    =
    \ap{T, \Phi D_{\mu} \psi}
    \\
    =
    \ap{D^{-\mu }D^\mu T, \Phi D_{\mu} \psi}
    =
    \ap{D^\mu T, D_{-\mu} \Phi D_{\mu} \psi}.
  \end{multline*}
  Taking into account that $D^\mu T$ has support contained in $(R,
  \infty)$ and that $D_{-\mu} \Phi D_{\mu} \psi$ is smooth and has
  support contained in $(-\infty, b)$, we have
  \begin{multline*}
    \abs{ \ap{D^\mu (\Phi T), \psi} }
    \leq
    \norm{ D^\mu T }_{L^1(R,b)}
    \norm{ D_{-\mu} \Phi D_{\mu} \psi }_{L^\infty(R,b)}
    \\
    \leq
    K
    \norm{ D^\mu T }_{L^1(R,b)}
    \norm{\psi}_\infty,
  \end{multline*}
  thanks to lemma~\ref{lem:bounded_operator}.
\end{proof}

\bibliographystyle{abbrv}
\bibliography{ozarfreo-latin1}

\begin{thebibliography}{10}

\bibitem{A99}
D.~J. Aldous.
\newblock Deterministic and stochastic models for coalescence (aggregation,
  coagulation): a review of the mean-field theory for probabilists.
\newblock {\em Bernouilli}, 5:3--48, 1999.

\bibitem{citeulike:2298278}
S.~Cueille and C.~Sire.
\newblock Nontrivial polydispersity exponents in aggregation models, May 1997.

\bibitem{D72}
R.~L. Drake.
\newblock A general mathematical survey of the coagulation equation.
\newblock In G.~M. Hidy and J.~R. Brock, editors, {\em Topics in Current
  Aerosol Research (Part 2)}, volume~3 of {\em International Reviews in Aerosol
  Physics and Chemistry}, pages 201--376. Pergamon, 1972.

\bibitem{citeulike:1069567}
M.~Escobedo and S.~Mischler.
\newblock Dust and self-similarity for the smoluchowski coagulation equation.
\newblock {\em Annales de l'Institut Henri Poincare (C) Non Linear Analysis},
  23(3):331--362, 2006.

\bibitem{MR2114413}
M.~Escobedo, S.~Mischler, and R.~M. Ricard.
\newblock On self-similarity and stationary problem for fragmentation and
  coagulation models.
\newblock {\em Ann. Inst. H. Poincar\'e Anal. Non Lin\'eaire}, 22(1):99--125,
  2005.

\bibitem{citeulike:1172261}
N.~Fournier and P.~Laurençot.
\newblock Existence of self-similar solutions to smoluchowski's coagulation
  equation.
\newblock {\em Communications in Mathematical Physics}, 256(3):589--609, June
  2005.

\bibitem{citeulike:1271620}
N.~Fournier and P.~Laurençot.
\newblock Local properties of self-similar solutions to {S}moluchowski's
  coagulation equation with sum kernels.
\newblock {\em Proc. Roy. Soc. Edinburgh Sect. A}, 136(3):485--508, 2006.

\bibitem{citeulike:2364173}
P.~Hartman.
\newblock {\em Ordinary Differential Equations (Classics in Applied
  Mathematics)}.
\newblock {Society for Industrial Mathematics}, March 2002.

\bibitem{citeulike:1103487}
M.~Kreer and O.~Penrose.
\newblock Proof of dynamical scaling in {S}moluchowski's coagulation equation
  with constant kernel.
\newblock {\em Journal of Statistical Physics}, 75(3-4):389--407, May 1994.

\bibitem{MR2068589}
P.~Laurençot and S.~Mischler.
\newblock On coalescence equations and related models.
\newblock In {\em Modeling and computational methods for kinetic equations},
  Model. Simul. Sci. Eng. Technol., pages 321--356. Birkh\"{a}user Boston,
  Boston, MA, 2004.

\bibitem{citeulike:1134194}
F.~Leyvraz.
\newblock Scaling theory and exactly solved models in the kinetics of
  irreversible aggregation.
\newblock {\em Physics Reports}, 383(2-3):95--212, August 2003.

\bibitem{citeulike:1103441}
G.~Menon and R.~L. Pego.
\newblock Approach to self-similarity in {S}moluchowski's coagulation
  equations.
\newblock {\em Communications on Pure and Applied Mathematics},
  57(9):1197--1232, 2004.

\bibitem{MR2139564}
G.~Menon and R.~L. Pego.
\newblock Dynamical scaling in {S}moluchowski's coagulation equations: uniform
  convergence.
\newblock {\em SIAM J. Math. Anal.}, 36(5), 2005.

\bibitem{citeulike:1428685}
K.~B. Oldham and J.~Spanier.
\newblock {\em The Fractional Calculus; Theory and Applications of
  Differentiation and Integration to Arbitrary Order (Mathematics in Science
  and Engineering, V)}.
\newblock {Academic Pr}, September 1974.

\bibitem{citeulike:1428686}
S.~G. Samko, A.~A. Kilbas, and O.~I. Marichev.
\newblock {\em Fractional Integrals and Derivatives: Theory and Applications}.
\newblock CRC, December 1993.

\bibitem{citeulike:1915345}
L.~Schwartz.
\newblock {\em Th\'{e}orie des distributions}.
\newblock Hermann, October 1997.

\bibitem{citeulike:1300131}
P.~G.~J. van Dongen and M.~H. Ernst.
\newblock Scaling solutions of smoluchowski's coagulation equation.
\newblock {\em Journal of Statistical Physics}, 50(1):295--329, January 1988.

\end{thebibliography}

\end{document}